\documentclass[a4paper,11pt]{article}
\usepackage[T1]{fontenc}
\usepackage[utf8]{inputenc}
\usepackage[english]{babel}
 
\usepackage{amsmath}
\usepackage{mathtools}
\usepackage{amssymb}
\usepackage{amsthm}
\usepackage{enumerate}
\usepackage{appendix}
\usepackage{geometry}
\usepackage[hyphens]{url}

\usepackage[colorlinks=true,linkcolor=blue,urlcolor=blue]{hyperref}
\usepackage{eurosym}
\usepackage{hyperref}
\hypersetup{
    colorlinks,
    citecolor=blue,
    filecolor=blue,
    linkcolor=blue,
    urlcolor=blue
}
\usepackage[hyphenbreaks]{breakurl}
\usepackage{breakcites}

 \usepackage{rotating}
 \usepackage[round]{natbib}

\usepackage{import}
\def\be{\begin{eqnarray}}
\def\ee{\end{eqnarray}}

\def\b*{\begin{eqnarray*}}
\def\e*{\end{eqnarray*}}

\newtheorem{Theorem}{Theorem}[part]
\newtheorem{Definition}{Definition}[part]
\newtheorem{Proposition}{Proposition}[part]

\newtheorem{Remark}{Remark}[part]

\newcommand{\ba}{\begin{array}}
\newcommand{\ea}{\end{array}}
\newcommand{\ben}{\begin{equation*}} 
\newcommand{\een}{\end{equation*}}
\newcommand{\bea}{\begin{eqnarray}}
 \newcommand{\eea}{\end{eqnarray}}
\newcommand{\bean}{\begin{eqnarray*}} 
\newcommand{\eean}{\end{eqnarray*}}
\newcommand{\bel}{\begin{align}} 
\newcommand{\eel}{\end{align}}
\newcommand{\beln}{\begin{align*}} 
\newcommand{\eeln}{\end{align*}}
\newcommand{\bit}{\begin{itemize}}
\newcommand{\eit}{\end{itemize}}

\makeatletter \@addtoreset{equation}{section}

\@addtoreset{Definition}{section}

\@addtoreset{Theorem}{section}

\@addtoreset{Proposition}{section}

\@addtoreset{Property}{section}

\@addtoreset{Assumption}{section}

\@addtoreset{Corollary}{section}

\@addtoreset{Lemma}{section}

\@addtoreset{Remark}{section}

\@addtoreset{Example}{section}

\@addtoreset{Condition}{section}

\def \E{\mathbb{E}}

\def \H{\mathbb{H}}
\def \L{\mathbb{L}}
\def \M{\mathbb{M}}
\def \N{\mathbb{N}}
\def \P{\mathbb{P}}
\def \Q{\mathbb{Q}}
\def \R{\mathbb{R}}

\def \Z{\mathbb{Z}}

\def \G{\mathbb{G}}

\def\Ec{{\cal E}}

\def\Hc{{\cal H}}

\def\Pc{{\cal P}}
\def\Tr#1{{\rm Tr}\left[#1\right]}

\def\={\;=\;}
\def\.{\;.}

\def\eps{\eps}

\def\1{{\bf 1}}

\def\eps{\epsilon}

 \def\normeL2#1{\left\|{#1}\right\|_{L^2}}
 
\newcommand{\alias}[2]{
\providecommand{#1}{}
\renewcommand{#1}{#2}
}

\alias{\P}{\mathbb{P}}
\alias{\N}{\mathcal{N}}
\alias{\L}{\mathcal{L}}
\alias{\Z}{\mathbb{Z}}
\alias{\Q}{\mathbb{Q}}
\alias{\R}{\mathbb{R}}
\alias{\C}{\mathcal{C}}
\alias{\T}{\mathbb{T}}
\alias{\E}{\mathbb{E}}
\alias{\H}{\mathcal{H}}
\alias{\1}{\mathbf{1}}
\alias{\B}{\mathcal{B}}
\alias{\M}{\mathcal{M}}
\alias{\G}{\mathcal{G}}
\alias{\Y}{Y_{\bullet}}

\def\Esp{\mathbb{E}}

\newcommand{\nc}{\newcommand}
\nc{\cA}{{\mathcal A}} \nc{\cB}{{\mathcal B}} \nc{\cC}{{\mathcal
C}} \nc{\cD}{{\mathcal D}} \nc{\bbD}{\mathbb{D}}
\nc{\cG}{{\mathcal G}} \nc{\cF}{{\mathcal F}} \nc{\cS}{{\mathcal
S}} \nc{\cU}{{\mathcal U}} \nc{\cH}{{\mathcal H}}
\nc{\cK}{{\mathcal K}} \nc{\cM}{{\mathcal M}} \nc{\cO}{{\mathcal
O}} \nc{\cP}{{\mathcal P}} \nc{\bbE}{\mathbb{E}}
\nc{\bbEP}{\mathbb{E}_{\mathbb{P}}}\nc{\bbL}{\mathbb{L}}
\nc{\bbP}{\mathbb{P}} \nc{\bbQ}{\mathbb{Q}} \nc{\del}{\partial}
\nc{\Om}{\Omega} \nc{\om}{\omega} \nc{\bbR}{\mathbb{R}}
\nc{\bbC}{\mathbb{C}} 
\nc{\dXt}{\delta q_{t}}
\nc{\dXs}{\delta q_{s}} \nc{\bs}{\blacksquare} \nc{\dX}{\delta q}
\nc{\dY}{\Delta Y}
\nc{\dnkx}{\left(X(T^{n}_{k})-X(T^{n}_{k-1})\right)}
\nc{\esssup}{\mathrm{ess}\mbox{ }\mathrm{sup}}
\nc{\essinf}{\mathrm{ess}\mbox{ } \mathrm{inf}}
\nc{\dhats}{\widehat{\delta_s}}

\nc{\chf}{\mbox{$\mathbf1$}}
\nc{\ind}{\mathds{1}}
\def\Esp{\mathbb{E}}

\begingroup\expandafter\expandafter\expandafter\endgroup
\expandafter\ifx\csname pdfsuppresswarningpagegroup\endcsname\relax
\else
\fi
\graphicspath{{./fig/}}

\nc{\mum}{ \mu_{\rm m} }
\nc{\muv}{ \mu_{\rm v} }
\nc{\mumv}{ \mu_{\rm mv} }
\nc{\Hm}{ H_{\rm m} }
\nc{\Hv}{ H_{\rm v} }

\newcommand{\ak}[1]{{\textcolor{black}{ #1}}}
 
\newcommand{\hs}{\vspace{3mm}}

\newcommand{\dd}{\mathrm{d}}

\usepackage[onehalfspacing]{setspace}
\newcommand{\uproman}[1]{(\uppercase\expandafter{\romannumeral#1})}

\usepackage{bbm}

\usepackage{graphicx}
\usepackage{float}
\graphicspath{{Figures_29-Dec-2023/}}
\usepackage{booktabs}
\usepackage{footnote}

\usepackage{caption}
\usepackage[skip=10pt]{subcaption}
\usepackage{bm}

 \nc{\HcA}{\mathcal{H}_{\rm A}}
 \nc{\HcP}{\mathcal{H}_{\rm P}}

\nc{\etai}{\eta_{ i}}
\nc{\etaj}{\eta_{ j}}
\nc{\etaA}{\eta_{\rm A}}
\nc{\etaP}{\eta_{\rm P}}
\nc{\asbm}{a^{{\rm sb,m}}}
\nc{\asbcn}{a^{{\rm sb,c}}_n}
\nc{\asbc}{a^{{\rm sb,c}}}
\nc{\vsbm}{v^{{\rm sb,m}}}
\nc{\vsbc}{v^{{\rm sb,c}}}
\nc{\Qm}{Q_{\rm m}}
\nc{\Qc}{Q_{\rm c}}
\nc{\bQ}{{\bold Q}}

\nc{\ac}{a^{{\rm c}}}

\nc{\Lm}{L^{\rm m}}
\nc{\Lc}{L^{\rm c}}
\nc{\tp}{\intercal}

\nc{\zm}{z^{\rm m}}
\nc{\zc}{z^{\rm c}}
\nc{\gm}{\gamma^{\rm m}}
\nc{\gc}{\gamma^{\rm c}}

\nc{\Am}{A^{\rm m}}
\nc{\Ac}{A^{\rm c}}
\nc{\Mm}{M^{\rm m}}
\nc{\Mc}{M^{\rm c}}
\nc{\Nm}{N^{\rm m}}
\nc{\Nc}{N^{\rm c}}

\nc{\pim}{\pi^{\rm m}}
\nc{\rim}{r^{\rm m}}
\nc{\rimi}{r^{\rm m}_i}
\nc{\rimj}{r^{\rm m}_j}

\nc{\pic}{\pi^{\rm c}}
\nc{\ric}{r^{\rm c}}
\nc{\ricii}{r^{\rm c}_{i,i}}
\nc{\ricij}{r^{\rm c}_{i,j}}

\nc{\bzeta}{{\bold \zeta}}

\nc{\xim}{\xi^{\rm m}}
\nc{\ximv}{\xi^{\rm m,v}}
\nc{\ximf}{\xi^{\rm m,f}}

\nc{\xic}{\xi^{\rm c}}
\nc{\xicv}{\xi^{\rm c,v}}
\nc{\xicf}{\xi^{\rm c,f}}

\begin{document}
\sloppy

\title{A Principal-Agent Model \\ for Optimal Incentives in Renewable Investments}

\author{René Aïd\thanks{Université Paris-Dauphine -- PSL Research University and Africa Business School, UM6P. Work supported by the {\em Finance for Energy Markets} Research Initiative,  the {\em Finance and Sustainable Development EDF-CA CIB Chair}, the EcoREES ANR-19-CE05-0042 project and the ANR PEPR Math-Vives project MIRTE ANR-23-EXMA-0011.}  \quad Annika Kemper\thanks{Center for Mathematical Economics (IMW) at Bielefeld University.
Financial support from the Deutsche Forschungsgemeinschaft (DFG, German Research Foundation) – SFB 1283/2 2021 – 317210226, DAAD and BGTS is gratefully acknowledged.} \quad Nizar Touzi\thanks{New York University, Tandon school of engineering. This work benefits from the financial support of the Chairs {\it Financial Risk}, and {\it Finance and Sustainable Development}.}}

\maketitle

\begin{abstract}
We investigate the optimal regulation of energy production in alignment with the long-term goals of the Paris Climate Agreement. We analyze the optimal regulatory incentives to foster the development of non-emissive electricity generation when the demand for power is met either by a single firm or by two interacting agents. The regulator aims to encourage green investments to limit carbon emissions while simultaneously reducing the intermittency of total energy production. We find that the regulator can achieve a higher certainty equivalent by regulating two interacting firms, each investing in one technology, rather than a single firm managing both technologies. This higher value is achieved thanks to a greater degree of freedom in the incentive mechanisms, which involve cross-subsidies between firms. Moreover, we find that it is optimal to compensate firms for shutting down their emissive production assets. We provide closed-form expressions of the second-best contracts and show that they take a rebate form, involving time-dependent prices for each state variable. A numerical study quantifies the impact of the designed second-best contract in both market structures compared to the business-as-usual scenario.
\end{abstract}

\noindent
\textit{JEL classification:}
C72, D86, Q28.

\noindent
\textit{Keywords:} 
Principal-Agent Problem,
Contract Theory, Moral Hazard, 
Extended Linear Quadratic Cost, 
Optimal Regulation,  
Green Investments,
Renewable Energy.

%\clearpage
%\tableofcontents 

%
% Introduction
%

\section{Introduction}

This paper investigates Principal-Multi-Agent incentive problems inspired by the need for a suitable `green' regulation reflecting long-term goals of the Paris climate agreement (see \cite{PCA}). Investments in renewable energies such as wind and solar energy play an important role in the current  climate debate. Due to their high fluctuations, however, conventional energy is still an attractive alternative for power producers even causing high carbon emissions.
Regulation of the market in line with the green deal is necessary such that renewable energy generation is enforced  to limit global warming.

However, an important feature of electricity generated by renewable energy, like solar and wind energy, is its {\em intermittency}. Their production depends on the realization of solar radiation and wind. Hence, the development of renewable energy increases the volatility of electricity production. Counter measures have to be taken into account to maintain the reliability of the power system. Thus, renewable energy provides both a positive externality thanks to the carbon emission they allow to avoid and a negative externality because of the indirect cost induced by their intermittency (see for an introduction to that economic literature \cite{Joskow11}, \cite{Borenstein12}, \cite{Hirth13} and \cite{Gowrisankaran16}).
In this context, it makes perfect sense for the regulator to provide  incentives to invest in renewables and also in counter measures to reduce the volatility they induce in the system (like storage or demand response enrollment programs).

Besides, the climate change and energy transition context, the present energy crisis that has taken over Europe in 2022 has triggered a series of public actions which trends toward an increase of the control of energy markets by the States.  Indeed, the crisis raised voices on the necessity of a new market regulation (see \cite{ReformElectricityMarket2022}) while in France, the main electricity utility holding more than 75\% of production capacity (in 2020) is becoming a fully national company.

This motivates our Principal-Agents approach to assess the optimal incentives mechanism to achieve an appropriate level of investment in non-emissive electricity production technology while maintaining smoothness of the energy production. Our approach builds on \cite{Cvitanic2018} which itself is based on the work by \cite{Sannikov2008}. In this work, \cite{Cvitanic2018} provides a general solution for the design in continuous-time of an optimal contract with moral hazard when one is concerned with both drift {\em and} volatility control of the state variable. This result is of particular interest in a context where reducing the volatility of the intermittent renewable energy production is of the utmost importance.  Their result was extended by \cite{EliePossamai2019} to the case of Principal-Multi-Agent models and by \cite{ElieMastroliaPossamai2019} in the case of a mean-field of Agents but restricted to drift control. This extension allows to deal with different polar market structures, like one regulated firm investing in all available technologies or a several firms investing each in one technology. Moreover, \cite{Hubert2023} develops a general Principal-Multi-Agent problem within hierarchical framework with drift and volatility. 

\hs

In this paper, we design an optimal contract, offered by the regulator (the Principal) encouraging firms (the Agents) to invest in renewable energy technology while stabilizing the energy production and disinvest in already existing emissive installed production capacities. In our model, the firms can invest and disinvest in two types of technology: intermittent renewable energy production  like wind farms and solar panels, or conventional production technologies which are not intermittent but emissive like gas or coal--fired plants. Agents receive a fixed proportional price for their energy production while facing linear quadratic investment costs.  We also introduce a small congestion cost term ($\eps$) proportional to the product of both investment rates to capture the idea that investing in both technology at the same time may result in congestion in the procurement of raw material or natural ressources like land. Further, they face a nonlinear volatility reduction cost function. Besides, installed capacities are prone to depreciation.   The energy producers are incentivised  to manage both the production level  of their generation assets and the volatility of their production. The regulator pays the producers for the energy they produce and bears three externalities. First, the conventional technology generates carbon emissions with constant unit value ($k_1$ in \euro/MWh), second, the renewable technology provides a positive social value of avoided emissions with a constant unit value ($k_2$ in \euro/MWh), and third, the renewable energy induces costly quadratic variation ($h$ in \euro/MW$^2$h). However, if the regulator can observe the amount of energy produced by each technology, she cannot observe the related efforts. Neither she can observe the efforts undertook by the Agents regarding the intermittency reduction of their production. These features lead to an incentive problem of moral hazard type. 

In particular, we consider two types of market structures. First, the electricity production is served by a single regulated firm which can invest in both technologies. Second, the electricity is served by two firms in interaction. Each firm can invest in only one technology: conventional technology for the first one and the renewable technology for the second one. This restriction on the space of potential technology, in which firms invest, reflects the existence of renewable only electricity production firms, like NextEra\footnote{As of September 19$^{\text{th}}$, 2022 NextEra wind and solar energy producer capitalisation is USD 168 billions while Exxon oil company is USD 388 billions.}. This setting allows to drive conclusions on the effect of interaction on the incentive mechanism required to achieve the appropriate level of renewable energy investments.  

As a surprise, we find that the regulation of two firms in interaction result in a higher certainty equivalent for the regulator compared to the case of a single firm handling the two technologies. The two market structures offer the same level of efficiency only when firms are risk-neutral and there is no congestion cost. In the other cases, the difference is in favor of the  market structure with interaction.  This purely economic advantage of the interactive market over the regulated single firm is due to a higher flexibility of the incentives and the possibility of designing cross-subsidies between firms. Hence, this gain also comes at the cost of a higher complexity of the incentive mechanism. Indeed, we say that the optimal incentives are coupled when the optimal incentive provision for one technology does depend on the investment cost parameters of the other technology and/or the risk-aversion parameter of the other firm in the case of interaction. 

In the case of a single firm, if the firm is risk-neutral or if there is no congestion cost  (i.e. $\eps=0$), the incentive provisions per technology are decoupled. In each case, the prices per technology basically boil down to the externality values they represent for the regulator. But, apart from these two extreme cases, the optimal incentives for the energy produced and for the energy production volatility reduction are fully coupled: prices paid for technologies depend on each other cost structure.

In the case of a two interacting firms, unless both firms are risk-neutral and there is no congestion cost, the optimal incentives per firm are fully coupled. For instance, the volatility reduction cost of the conventional energy has an effect on the payment rate of renewable energy production. This coupling does not burden much the computation of the incentive prices. But, it may induce an acceptance issue from the market participants  by making the incentive mechanism of each firm interdependent.

\hs

The recent literature already considers optimal installation of renewable power plants (see \cite{KochVargiolu2021}, \cite{AwerkinVargiolu2021}) as a stochastic control problem of a price--maker company. Besides,  \cite{Kharroubi2019} focus on the regulation of renewable resource exploitation  assuming a geometric Brownian motion for the state dynamics under drift control only.  However, as pointed out by \cite{FloraTankov2022} under the current policy scenario, the trend evolution of green house gas emissions reduces insufficiently. In order to reach the net zero 2050 goal, further regulatory action seems unavoidable. Further works already consider governmental incentives for green bonds investment (see \cite{BaldacciPossamai2022}) or for lower and more stable energy consumption (see \cite{Aid2022} and \cite{Elie2020}). However, it seems also natural to directly encourage energy producers in investing into a more `green' energy production by regulatory action within a Principal-(Multi-)Agent framework. 

Note that there is also an extensive recent literature on mean-field games with applications to the energy sector, such as \cite{Carmona21} on the regulation of carbon emissions in energy production, by \cite{Aid20} on distributed electricity generation, by \cite{Alasseur22} and \cite{Firoozi22} on energy certificates, and by \cite{AlasseurBaseiBertucciCecchin2023} for a model with a mean-field of competitive producers of renewable energies and its comparison with a monopoly.

\hs

The paper is organized as follows:  Section~\ref{sec:model} describes our model of the development of renewable energy production in two specific market structures, a single energy producer and an interactive setting with two energy producers. Section~\ref{sec:results} provides the main results of this paper while Sections~\ref{sec:oneprod} and \ref{sec:twoprod} give the detailed solutions of the two incentive problems. Section \ref{sec:Application_Numerics}  illustrates numerically the impact of renewable regulation. Finally, Section \ref{sec:Conclusion} concludes.

%
% Main Model
%
\section{Market structure for renewable energy development}\label{sec:model}

We present here the common features of our model of incentives for development of renewable energy where production is served by a single producer or by two competing firms.

\subsection{Linear controlled dynamics for conventional and renewable energy}

The controlled state process is denoted by $X=(X_t)_{t\in[0,T]}$ and is the canonical process of the space~$\Omega$ of $2-$dimensional continuous trajectories $\omega\colon[0,T]\to\mathbb{R}^2$, i.e. $X_t(\omega)=\omega(t)$ for all $(t,\omega)\in[0,T]\times\Omega$. We shall denote by $\mathbb{F}=\{\mathcal{F}_t,t\in[0,T]\}$ the corresponding filtration with appropriate completion as in \cite{Aid2022}. 

Throughout this paper we denote by $\Pc$ the collection of all probability measures $\P$ such that $X_0=(X^1_0,X^2_0)\in\mathbb{R}^2$ is given and
\begin{align}\label{X}
 \dd X_t^i=\left(a_t^i-\delta_i X_t^i\right) \dd t+b_t^{i,1}\dd W_t^{i,1} + b_t^{i,2}\dd W_t^{i,2}\;, \quad i=1,2\;,~\P-\mbox{a.s.}
\end{align}
for some independent $\P-$Brownian motions $W^{\P}=(W^{i,j})_{1\le i,j\le 2}$ and some $\R\times\R^2-$progressively measurable processes $\nu^i:=(a^i,b^i)$ with 
\b*
b^i_t\neq 0
&\mbox{for all}~t\in[0,T],\,\, i=1,2,~\mbox{and}&
\E^{\P}\big[\Ec(L^{\nu^1}\!\!+\!L^{\nu^2})_T\big]
=1,~~
L^{\nu^i}_t
:=
-\int_0^t\frac{a^{i}_tb^{i}_t}{|b^{i}_t|^{2}}\cdot \dd W_t^{i},
\e*
and $\Ec$ denotes the Dol\'eans-Dade exponential.
Here, $\P$ is the distribution of the state process $X$, and the corresponding process $\nu^\P=(\nu^1,\nu^2)$ represents the effort on conventional emissive energy production $X^1$ and the renewable technology energy production $X^2$. The coefficients $\delta_1,\delta_2\ge 0$ are depreciation rates for the two technologies which might be different so as to capture different lifespan for the two production capacities (e.g. $40$ years for coal-fired power plants versus $25$ years solar panels or wind farms). \footnote{Our model restrict sources of shocks to those generated by Brownian motions. We may extend to Poisson process shocks so as to cover dynamics of outages of power plants for instance, see e.g  \cite{Euch21} in the context of platform mechanism design modeling.}

Let us comment on the dynamics \eqref{X} of the energy production system.
\begin{itemize}
\item The independence between the sources of randomness $W^{1,.}$ and $W^{2,.}$ affecting the two energy productions $X^1$ and $X^2$ can be justified by the fact that renewable production assets like solar and wind are highly affected by  weather conditions while coal and gaz fired plants are mostly affected by independent outages causes. 
\item Moreover, the dynamics \eqref{X} models each technology by means of a controlled stochastic differential equation affected by two sources of shocks. Indeed, this feature captures the idea that the aggregate production of a technology can be affected by several random factors. In the case of wind production, one can think of different sources of wind depending on the locations; in the case of gas or coal fired plants, differents sources of outages. We limited these sources of random shocks to two for sake of computational burden. Moreover, production dynamics is subject to two sources of noise with different corresponding costs for the effort which are not observable through the observation of $X$. Instead only $\frac{\dd}{\dd t}\langle X^j\rangle_t = |b^{j}_t|^2$ is observable, inducing an information asymmetry at the level of the volatility effort in addition to the standard one on the drift effort $a^i$. This point will guarantee that the regulator cannot infer the level of efforts of the firms on each sources of randomness. 
\end{itemize}

\subsection{Separable cost of effort}

Next, we introduce a cost function which is separable in the action variables $a$ and $b$:
\begin{align}\label{C}
 C(a,b)= C_1(a,b)+C_2(a,b),
 ~\mbox{where}~
 C_i(a,b):=g_i(a)+ \phi_i(b^i),
\end{align}
with linear--quadratic investment cost functions:
\begin{align}\label{g}
&   g_i(a) =  l_ia_i + \frac12 q_i a_i^2+ \frac12 \eps a_1a_2,
   ~i=1,2,
   ~\mbox{with the condition}~
   \Qm :=q_1q_2-\eps^2>0,
\end{align}
for some given parameters $l_1,l_2,q_1,q_2>0$, and a separable cost function for the action on production volatilities:
\begin{align}\label{eq:VolaCostExplicit_M}
\phi_i(b_i) := \phi_{i,1}(b_{i,1})\!+\!\phi_{i,2}(b_{i,2}),
~
\text{where}~
\phi_{i,j}(b_{i,j}) := \left( b_{i,j}^{-2}\!-\!\sigma_{i,j}^{-2}\right)\Phi_{i,j},
~b_{i,j} \in (0,\sigma_{i,j}],
\end{align}
for some given parameters $\Phi_{i,j},\sigma_{i,j}>0$, $i,j=1,2$.\footnote{Similar to \cite{Aid2022}, we should also have imposed that the efforts $b^{i,j}$ are uniformly bounded from below above zero. We refrain from introducing this additional technical aspect for simplicity and we emphasize that it can be handled by following the same line of argument as in \cite{Aid2022}.}

The condition $Q_{\rm m}>0$ is necessary and sufficient for the map $a\longmapsto (g_1+g_2)(a)$ to be strictly convex. We also notice that we may rewrite the drift effort cost function as $g_i(a)=[l_i \frac12 \eps(a_1+a_2)]a_i + \frac12 (q_i-\eps)a_i^2$, thus interpreting the cross effort term as a dependence of the marginal cost of effort on the total effort exerted on both energy productions. 

The investment cost functions $g_j$ consist of three components. Notice that the action variable $a_j$ has no sign restriction meaning that the firm may decide to remove some production capacity. For small values of $a$, the cost is essentially driven by the linear part $l_ja_j$ implying that removal of production capacities allows to recover the corresponding cost. However, due to the quadratic component, this does not hold for general values of $a$. Hence small investments are reversible at small cost. The interaction component of cost $\frac12 \eps a_1a_2$ models the competition between technologies for some ressources (raw material, land,...). Investing in both technologies induces a congestion cost in ressource provision. Similarly, disinvesting in one technology while investing in the other leads to an overall cost reduction. This would be the case for instance if the firm recycles the materials from gas-fired plants to build wind farms. Note that if $\eps = 0$, the investment problem of the firm is decoupled in two distinct problems.

To illustrate this separability limitation, we may think of the energy production of a wind farm  affected by the wind and by outages whose management costs are separable in the following sense: the wind source of randomness may be mitigated by using batteries, while outages randomness may be managed by appropriate default forecasting tools. In both cases, when the firm takes no action i.e $b_{j,l}=\sigma_{j,l}$, it bears no cost and the baseline volatility of the shocks is $\sigma_{j,l}$. When the firm undertakes some effort i.e. $b_{j,l}<\sigma_{j,l}$ the volatility is reduced at a cost which goes to infinity when $b_{j,l}$ goes to zero. Thus, the production uncertainty cannot be reduced to zero at finite cost.

\subsection{Regulator facing one producer}

The execution of the contract begins at $t=0$. We start with the one single producer case. The production firm sells the total production of energy at the unit price $p\in\mathbb{R}$ and undergoes the cost of effort. In addition, the production firm receives a lump sum incentive compensation $\xi$ from the regulator at maturity $T$, where negative payment indicates a charge. The objective function of the firm is then defined by
\begin{align}
J^{\rm m}_{\rm A}(\xi,\mathbb{P})
:= 
\Esp^{\P}\Big[U_{\rm A}\Big(\xi \!+\!\!\int_{0}^{T}\!\!\!\! \big\{p (X^1_t\!+\!X_t^2) \!-\! C(a^\P_t,b^\P_t)\big\}\dd t\Big)\Big],
~\text{with}~U_{\rm A}(x) = -e^{-\eta_{\rm A} x},
\end{align} 
for some constant risk aversion parameter $\eta_{\rm A}>0$. Here, the superscript ${}^{{\rm m}}$ refers to the monopolistic setting of this section.\footnote{The ``monopolistic'' denomination might be misleading as it should also come with a different equilibrium price $p$ which is not considered in our model. Nevertheless, it represents an helpful mnemonic name when comparing with the case of two interacting firms.} 
In the present setting, the investments in energy sources are perfectly substitutable as $X^1$ and $X^2$ have the same purpose for the firm. The problem of the single energy producer is
\begin{align}\label{eq:pb1firm}
V^{\rm m}_{\rm A}(\xi) = \sup_{\P\in\mathcal{P}}J_{\rm A}(\xi,\mathbb{P})\;.
\end{align}
A control $\widehat\P^{\rm m}\in \mathcal{P}$, with $\widehat\nu^{\rm m}:=\nu^{\widehat\P^{\rm m}}=(\widehat a^{\rm m},\widehat b^{\rm m})$  will be called optimal if $V^{\rm m}_{\rm A}(\xi)=J_{\rm A}(\xi,\widehat\P^{\rm m})$. We denote the collection of all such optimal responses by $\widehat{\mathcal{P}}^{\rm m}(\xi)$.

\hs

The case when the firm receives no incentive payment from the regulator $\xi \equiv 0$ will be referred to as the {\em business-as-usual} regime.

\hs

The regulator's objective function is given by
\begin{align*}
\!\!\!J^{\rm m}_{\rm P}(\xi,\P) 
:=
\Esp^{\P}\!\Big[U_{\rm P}\Big(\!\!-\xi 
\!+\! \int_{0}^{T} \!\!\!\!\big\{\!\!-\!p(X_t^1 \!+\! X_t^2) \!+\! k_1X_t^1 \!+\! k_2X_t^2\big\}\dd t
- \frac{h}{2}\langle X^1\!+\!X^2\rangle_T\Big)\Big],
\end{align*} 
where $U_{\rm P}(x)=-e^{-\eta_{\rm P} x}$, for some constant risk aversion parameter $\eta_{\rm P}>0$. This objective function contains four cost components. First, the regulator pays the incentive compensation $\xi$ defining contractual mechanisms that would still maintain the firm above some participation certainty equivalent utility $R_0^{\rm m}$. The regulator also pays the price for the total energy production, meaning that she has an interests in maintaining as low as possible the total electricity bill. Next, in contrast with the firm, the regulator takes also into account positive and negative externalities arising from the difference in technologies. She benefits from investments in renewables ($k_2>0$) and is harmed by carbon emissions resulting from conventional energy ($k_1<0$). 
Finally, the volatility of total electricity production, measured by its quadratic variation $\langle X^1\!+\! X^2\rangle$, represents a negative externality to the electric system with a constant marginal cost of $h>0$. The parameter $h$  is called the {\em direct cost of volatility}. Indeed, intermittency of production calls for costly adjustment of dispatchable production to maintain the balance of the system between production and consumption in the same way as described in \cite{Aid2022}. But, it should be noted that even in the absence of direct cost of volatility, the firm and the regulator are prone to make volatility reduction effort because of their risk--aversion. This indirect cost of volatility is referred to as the {\em risk-premium for volatility}. In the sequel, we rewrite the regulator's objective function as
\begin{align}\label{Jm}
\!\!\!J^{\rm m}_{\rm P}(\xi,\P) 
=
\Esp^{\P}\!\Big[U_{\rm P}\Big(\!\!-\xi 
\!+\! \int_{0}^{T} \!\!\!\!(k^p_1X_t^1 \!+\! k^p_2X_t^2) \dd t
- \frac{h}{2}\langle X^1\!+\!X^2\rangle_T\Big)\Big],
~~k^p_i:=k_i-p.
\end{align} 
The regulator acts as a leader anticipating the agent's optimal response. As standard in the principal-agent literature, we assume that, in case of multiple solutions in $\widehat{\mathcal{P}}^{\rm m}(\xi)$, the agent returns the most favorable response to the principal. The regulators's objective is then defined by the value function: 
\begin{align}
V_{\rm P}^{{\rm sb,m}}
=
\sup_{{\xi\in\Xi^{\rm m}}}~\sup_{\P\in\widehat{\mathcal{P}}^{\rm m}(\xi)} 
J^{\rm m}_{\rm P}(\xi,\P), 
~\mbox{where}~
\Xi^{\rm m}=\{\xi\in\mathcal{C}^{\rm m}\colon V_{\rm A}(\xi)\geq U_{\rm A}(R_0^{\rm m})\},
\label{eq:VF_Principal_SIATS}
\end{align}
for some participation certainty equivalent utility $R_0^{\rm m}$. Here, the superscript $^{\rm sb}$ stands for the second-best optimal contracting problem in the classical  Principal-Agent modeling, and $\mathcal{C}^{\rm m}$ is the set of all $\mathcal{F}_T-$measurable random variables, satisfying the integrability condition
\begin{align}
\sup_{\P\in\mathcal{P}}
\mathbb{E}^{\P}\left[e^{\mu\eta_{\rm P}  \xi }\right]
+ \sup_{\P\in\mathcal{P}}\mathbb{E}^{\P}\left[e^{-\mu\eta_{\rm A}  \xi}\right]<+\infty, 
\quad \text{for some } \mu>1.
\end{align}
Proposition \ref{prop:sbxi} (i) below provides an explicit expression of the optimal second-best contract by following the \cite{Sannikov2008}'s methodology, further justified and extended by \cite{Cvitanic2018}. This method relies on the following representation of admissible contracts in terms of a fixed payment $Y_0$ and pay-for-performance sensitivity processes $Z$ and $\Gamma$ as
\be\label{opt-contract}
\xi
=
Y_0+\int_0^T Z_t\cdot \dd X_t+\frac12{\rm Tr}\big[(\Gamma_t+\eta_{\rm A}Z_tZ_t^\intercal) \dd \langle X\rangle_t\big]-\mathcal{H}_{\rm A}(X_t,Z_t,\Gamma_t)\dd t,
\ee
where $\mathcal{H}_{\rm A}$ is the so-called Agent's Hamiltonian, see Section \ref{ssec:contract1} for the detailed analysis.

\subsection{Regulator facing two competing producers}

We now consider the case where each energy production $X^i$ is delegated to a separate production firm $i=1,2$. The first firm invests only in conventional emissive production capacity $X^1$ and chooses a probability measure which only affects the first component of the corresponding effort $\nu^{\P,1}=(a^1,b^1)$, while the second invests in renewable capacity $X^2$ and chooses a probability measure which only affects the second component of the effort process $\nu^{\P,2}=(a^2,b^2)$. Let
\b*
\Pc^1(\P)
:=
\big\{\P':\nu^{\P',2}=\nu^{\P,2}\big\}
~\mbox{and}~
\Pc^2(\P)
:=
\big\{\P':\nu^{\P',1}=\nu^{\P,1}\big\},
&\mbox{for all}&
\P\in\Pc.
\e*
Assuming that the agents preferences are characterized by exponential utilities $U_i(x)=-e^{-\etai x}$ with constant risk aversion parameter $\eta_i>0$, we introduce for all $\P\in\Pc$ the value function of each agent:
\begin{align}
V_i(\xi_i,\P)
:=\!\!
\sup_{\P'\in\mathcal{P}^i(\P)}\;
J_i(\xi_i,\P'), 
~\mbox{with}~
J_i(\xi_i,\P')
:= 
\mathbb{E}^{\P'}
\Big[U_i\Big(\xi_i+\!\int_{0}^{T}\!\!\!\big\{pX_t^i- C_i(\nu^{\P'}_t)\big\}\dd t \Big)\Big],
~ i=1,2.
\label{eq:VF_Agents_Competitive}
\end{align}
Here, the cost functions $C_i$ and $g_i$ have been introduced in \eqref{C}-\eqref{g}. We shall consider equilibrium situations between the two firms with individual optimization problem of Agent $i$ defined by:

\begin{Definition}[Production firms in Nash equilibrium]
A probability measure $\widehat\P$ corresponding to firms efforts $\widehat\nu:=\nu^{\widehat\P}=({\widehat\nu}^1,{\widehat\nu}^2)$ is a Nash equilibrium if $V_i(\xi_i,\widehat\P)=J_i(\xi_i,\widehat\P),~i=1,2$. 
\end{Definition}

In our setting, we shall see that there exists a unique Nash equilibrium $\widehat\P(\xi)$ induced by a unique $\widehat\nu(\xi)$ for all contract $\xi=(\xi_1,\xi_2)$.

Similar to \eqref{Jm}, we now introduce the objective function $J^{\rm c}_{\rm P}$ of the regulator, given the payments $\xi=(\xi_1,\xi_2)$ to the production firms:
\begin{align}\label{JPc}
J_{\rm P}^{\rm c}(\xi,\P)
:=
\mathbb{E}^{\P}\!
\Big[U_{\rm P}\Big(\!\!-\xi_1\!-\!\xi_2
                              +\!\!\int_{0}^{T}\!\!\!\!\big\{k^p_1X_t^1+k^p_2X_t^2\big\} \dd t
                              -\frac{h}{2}\langle X^1\!\!+\!X^2\rangle_T
                      \Big)
\Big],
\end{align} 
with $U_{\rm P}(x)=-e^{-\eta_{\rm P} x}$ for some constant risk aversion parameter $\eta_{\rm P}>0$, and where the superscript $^{\rm c}$ stands for the current competitive setting between the two production firms. Finally, the regulator's delegation problem if defined by:
\begin{align}
V_{\rm P}^{{\rm sb,c}}
:=
\sup_{\xi\in\Xi^{\rm c}}
J_{\rm P}^{\rm c}(\xi,\widehat\P(\xi)), 
~\mbox{with}~
\Xi^{\rm c}=\Big\{\xi\in\mathcal{C}^{\rm c}\colon V_i\big(\xi_i,\widehat\P(\xi)\big)\ge U_i(R_i^{\rm c}),  i=1,2\Big\},
\label{eq:VF_Principal_TIATS}
\end{align}
with given firms participation certainty equivalent utilities $R_1^{\rm c},R_2^{\rm c}$, and where $\mathcal{C}^{\rm c}$ is the set of all $\mathcal{F}_T-$measurable random variables, satisfying the integrability condition
\begin{align}
    \sup_{\P\in\Pc}
    \mathbb{E}^{\P}\Big[e^{\mu\eta_{\rm P} (\xi_1+\xi_2)}\Big]
    + \sum_{i=1}^{2}\sup_{\P\in\Pc}\mathbb{E}^{\P}\left[e^{-\mu\eta_i  \xi_i}\right]<+\infty, \quad \text{for some } \mu>1.
\end{align}
Unfortunately, the Sannikov's (2008) approach has not been extended in the literature to the current situation with controlled effort on the volatility. For this reason, our main results below will instead characterize the lower bound $\underline{V}_{\rm P}^{{\rm sb,c}}$ where the agent's contracts $\xi_i$ are restricted to be of the form \eqref{opt-contract} with the corresponding Hamiltonian $\Hc_{i}$ as defined in Section \ref{sec:Hi} below:
\begin{align}
V_{\rm P}^{{\rm sb,c}}
\;\ge\;
\underline{V}_{\rm P}^{{\rm sb,c}}
\;:=\;
\sup_{\xi\in\underline{\Xi}^{\rm c}}
J_{\rm P}^{\rm c}(\xi,\widehat\P(\xi)), 
~\mbox{with}~
\underline{\Xi}^{\rm c}=\Big\{\xi\in\Xi^{\rm c}: \xi_i~\mbox{of the form}~\eqref{opt-contract},  i=1,2\Big\},
\label{eq:underline-VF_Principal_TIATS}
\end{align}
Proposition \ref{prop:sbxi} (ii) below provides an explicit expression of the optimal second-best contract for the Principal's lower bound value $\underline{V}_{\rm P}^{{\rm sb,c}}$. In the case where the agents and the principal share the same risk aversion parameters, Theorem \ref{thm:comparison} below builds on the explicit results of Proposition \ref{prop:sbxi} to show that $\underline{V}_{\rm P}^{{\rm sb,c}}\ge V_{\rm P}^{{\rm sb,m}}$, which implies that the principal's second-best value function $V_{\rm P}^{{\rm sb,c}}$ under delegation to two competitive firms is larger than the corresponding value function $V_{\rm P}^{{\rm sb,m}}$ under delegation to one single integrated firm. 

\begin{Remark}
{\rm \begin{itemize}
\item[$(i)$] The setting above is not in line with the usual classical duopoly framework where the price of the commodity depends on the total amount produced. This constant price may be interpreted as a price regulated by another authority.
\item[$(ii)$] In our framework, firms are highly specialized as they only invest in one technology either emissive or renewable. New entrants like the company NextEra satisfy this specialization in renewable, but other utilities in real markets have diversified portfolio of technologies. Nevertheless, our objective is to design a simple model to assess the interaction between the investment in these two technologies and we leave for future research the analysis of the general case of investments in all technologies by all firms. 
\end{itemize}}
\end{Remark}

\section{Main results} \label{sec:results}
 
In Sections~\ref{sec:oneprod} and \ref{sec:twoprod}, we provide explicit solutions for optimal contracting problems by exploiting the linear-quadratic structure of the {\it monopolistic} and the {\it competitive} market structures respectively. In particular we show that the optimal contracts in both structures are of the rebate form. In the present section, we condensate the results on the optimal second-best contracts in both production organisation in the following proposition.

\begin{Proposition}{\rm (Second-best contracts)}\label{prop:sbxi}
\begin{enumerate}[\upshape (i)]

\item In the case of the regulation of a single firm, the second-best optimal incentive mechanism is given by 
$\xim = \ximf + \ximv_1 + \ximv_2$ where $\ximf$ is a constant payment and $\ximv_i$  $i=1,2$ are two variable payments given by
\begin{align}\label{eq:main:xim}
\ximv_i & = \int_{0}^{T} \pim_i(t)  (X_t^i-X_0^i) \dd t - \frac12 \int_{0}^{T}  \rimi(t) \dd \langle X^i_\cdot \rangle_t, \\
 \text{with} & \quad
\pim_i(t) = - \dot{z}^{\rm m}_i(t) - p + \delta_i z^{\rm m}_i(t),\quad
\rimi(t) =  h + \etaP \big(w_i(t) - z^{\rm m}_i(t)\big)^2 > 0, \label{eq:pisbm}
\end{align}
where $w_i,z^{\rm m}_i$, $i=1,2$ are deterministic functions of time given in \eqref{wi} and Proposition~\ref{prop:oneprodsb}, respectively.

\item In the case of the second-best regulation of two competing firms, the lower bound $\underline{V}_{\rm P}^{{\rm sb,c}}$ on the Principal's value function is attained by the optimal incentive scheme $\xic_i = \xicf_i + \xicv_{i,i} + \xicv_{i,j}$ for each firm $i$, for some constant payment $\xicf_i$ and running payments $\xicv_{i,i}$ and $\xicv_{i,j}$ given by
\begin{align}\label{eq:main:xic}
&\xicv_{i,j} = 
\int_{0}^{T} \pic_{i,j}(t) (X_t^j - X_0^j) \dd t 
- \frac12 \int_{0}^{T} \ricij(t) \dd \langle X^j\rangle_t 
+ (X_T^j - X_0^j) z_{i,j}(T), \quad j=1,2\\
 &
\pic_{i,i}(t)   = -\dot{z}^{\rm c}_{i,i}(t) - p + \delta_i \zc_{i,i}(t), \quad
\pic_{i,j}(t)  = -\dot{z}^{\rm c}_{i,j}(t)   + \delta_j \zc_{i,j}(t),~\mbox{for}~i\neq j, \label{eq:SBpricesDrift} \\
& \ricii(t) =  
h + \etaj \zc_{j,i}(t)^2+ \eta_{\rm P} \Big(\zc_{i,i}(t) + \zc_{j,i}(t) - w_i(t)\Big)^2,
\quad
\ricij(t)= -\etai \zc_{i,j}(t)^2,~\mbox{for}~i\neq j,
\end{align}
where $z^{\rm c}_{i,j}$ are deterministic functions of time given in Proposition~\ref{prop:twoprodsb}.

\end{enumerate}
\end{Proposition}

\hs

The functions $\pim_i, \pic_{i,j}$ and $\rim_i,\ric_{i,j}$ represent the second-best prices for the  energy and penalties for the intermittency, respectively. Indeed, $\pi^{\rm m}_i, \pi^{\rm c}_{i,j}$ is measured in \euro/MWh while $z^{\rm m}_i$ and $\zc_{i,j}$ are measured in \euro/MW. The dimensionality of the penalties for intermittency $r^{\rm m}_i,r^{\rm c}_{i,j}$ remains in \euro/MW$ ^2$.  Note that the form of the contract given by the expression \eqref{eq:main:xim} allows a direct implementation of the incentive mechanism. Indeed, the regulator only has to set time-varying deterministic prices in \euro/MWh for energy production $\pi^{\rm m}_j$ and penalties $\rim_j$ in \euro/MWh$^{2}$ for intermittency  to determine the payments and charges to provide to the firm. 

Moreover, we observe that the optimal incentive mechanism in the case of a single regulated firm has three terms. First, the firm gets a fixed payment $\ximf$ whose aim is to satisfy the participation constraint of the firm. Moreover, the firm receives or pays a variable term $\ximv_i$ depending on each technology  $i=1,2$. The variable terms admit a rebate form: the firm is paid at the end of the contracting period for what it realized in comparison to the initial state serving as a baseline.

\hs

The limiting case of zero depreciation rate, no technology cost interaction ($\eps=0$) and no volatility control allows to observe more clearly an interesting consequence of the form of the optimal mechanism~\eqref{eq:main:xim}. In that case, it holds that the prices for the energy are constant given by $\pim_j = (1- \eta_{\rm A} K )k_j - p$ where $K$ is a nonnegative constant (see remarks following Proposition~\ref{prop:oneprodsb}). The sign of the energy prices are driven by the value of the externality. For renewable energy, $k_2$ is positive and thus it is not surprising that the regulator pays the firm when $X^2_t- X^2_0$ is positive. But for an emissive technology, $k_1$ is negative: the regulator's mechanism implies to tax the firm for each MWh of emissive production but also {\em to pay the firm for each MWh of avoided emissive production}. In other words, electric utilities owning coal-fired plants should be paid when they shut them down.

\hs

We turn now to the situation of the regulation of two competing firms. We observe first that the optimal incentive mechanism of each firm exhibits some common features with the case of a single firm: the decomposition into a fixed part and a variable part with a rebate form. Even the expression of the prices $\pic_{i,i}$ for the energy have the same expression as in the single firm case. However, the mechanism presents several new terms. In particular,  firm~1 now receives a payment indexed on the firm~2's capacity trajectory. For example, emissive firm~1 receives a payment when the renewable capacity $X^2_t$ is larger than its initial condition. By this way, the emissive firm~1 has an incentive to see firm~2 responds appropriatly to the regulation mechanism. Moreover, even in the absence of covariation between $X^1$ and $X^2$ and in the absence of interaction in the controls of volatilities, firm~1 is compensated at the price $r^{\rm c}_{1,2}$ for the cost the firm~1 incurs because of firm~2's volatility. This is due to the fact that the incentive payment rate $\zc_{i,j}$ introduces a dependence of the firm~$i$ on the volatility of the technology~$j$, and thus firm~$i$ has to be compensated for this induced risk. Finally, a term for the terminal capacity is involved and might be non-zero depending on the value of $\zc_{i,j}(T)$. 

In the case of a two interacting firms, unless both firms are risk-neutral and there is no congestion cost, the optimal incentives per firm are fully coupled (see Proposition~\ref{prop:twoprodsb}~(i)). For instance, the volatility reduction cost function of the conventional energy has an effect on the payment rate of renewable energy production. This coupling does not burden much the computation of the incentive prices. But, it may induce an acceptance issue from the market participants  by making the incentive mechanism of each firm interdependent. 

\hs

As a by-product of our explicit calculations, we obtain the following remarkable observation that the regulator gets higher social value by delegating the energy production to two interacting firms rather than one single firm.
 
\begin{Theorem} {\rm (Market structure comparison)} \label{thm:comparison}
Assume $\eta_{\rm A} = \eta_1 = \eta_2$ and $R_0^{\rm m} \le R_1^{\rm c}  + R_2^{\rm c}$. Then, we have $V^{\rm sb,m}_{\rm P} \le V^{\rm sb,c}_{\rm P}$, i.e. the regulator's second-best value function when facing two competing agents is larger than the corresponding value when facing one single integrated agent.
\end{Theorem}

\noindent{\bf Proof}: See Appendix~\ref{app:compare}.

\hs

This result is surprising in the following sense. In the business-as-usual cases, it is clear that a single firm controlling the development of both technologies can generate more value than two firms each controlling one technology and having to coordinate. So the first intuition is that the regulation of these two market structures should maintain this inequality. But our result shows that, under optimal incentive mechanisms, the regulator can induce any effort level of the individual firm in the situation of two interacting firms, thanks to its larger number of incentive payment rates. 

The scope of this result should be tempered by the following limitations.  First, it only concerns the certainty equivalent of the regulator's value. It abstracts from firms' reservation utilities, which may be different and not comparable. Moreover, it assumes the same risk aversion parameter for all agents, which denies the possibility of having a lower risk aversion if the firm has the possibility to diversify its investments.  Second, the social benefit of regulating the two interacting firms is obtained at the cost of greater complexity. In this situation, the optimal incentive mechanism introduces cross payments between the firms., which may raise obstacles to its implementation.

\hs

Nevertheless, despite these limitations, to the best of our knowledge, this is the first result of its kind in the continuous-time optimal incentive literature. This result calls for a generalization, in particular, to the case where both firms can invest in both technologies. We leave for future research the investigation of its possible extension.

\section{Solution of the regulator facing one firm case} \label{sec:oneprod}
\subsection{The Contract}\label{ssec:contract1}

The Hamiltonian of the single firm problem is defined for all $x, z \in\mathbb{R}^2$,  $\gamma\in\mathcal{S}^{2}$ by:
 \begin{align}
 \HcA(x,z,\gamma) 
 =  
 \Hc^0(x,z) + \overline\Hc(z) +  \langle\Hc\rangle(\gamma),
 ~\mbox{with}~
 \Hc^0(x,z) = \sum_{j=1}^{2}(p-\delta_j z_j)x_j,
 \end{align}
 and nonlinear parts
 \begin{align}
 \overline\Hc(z) = \sup_{a} \sum_{j=1}^{2} \big\{z_j a_j  - g_j(a)\big\}, \quad
 \langle\Hc\rangle(\gamma) =  \sup_{b}\Big\{  \frac12 \Tr{b b^\top \gamma} - \phi(b)\Big\}.
 \end{align}
Following \cite{Sannikov2008} and \cite{Cvitanic2018} the regulator searches for contractual incentive mechanisms of the form $\xi = Y_T^{Y_0,Z, \Gamma}$ with
\begin{align}\label{eq:xi1}
Y_T^{Y_0,Z, \Gamma} \!\!= 
Y_0 \!+\!\! \int_{0}^{T}\!\! Z_t \!\cdot\! \dd X_t 
+ \frac12 \int_{0}^{T} \!\!\!\! \Tr{(\Gamma_t + \eta_{\rm A} Z_t Z_t^\top) \dd \langle X\rangle_t} - \int_{0}^{T} \!\!\!\!\HcA(X_t,Z_t,\Gamma_t)\dd t,
 \end{align}
for a given pair of processes $(Z, \Gamma)$ and some constant $Y_0$ with $U_A(Y_0)\ge U_{\rm A}(R_0^{{\rm m}})$. Here $Y$ is the continuation certainty equivalent utility of the producer, $Z$ represents the Sannikov's instantaneous payment rate for the production capacities $X$ and $\Gamma$ the instantaneous penalties for the corresponding risk induced by the quadratic variation $\langle X\rangle$.

The best response of the firm to instantaneous payments $z,\gamma$ is given by the maximizers of $\overline\Hc$ and $\langle\Hc\rangle$. Under the condition $Q_{\rm m}>0$ in \eqref{g} we derive the optimal response maps by means of the first order condition: 
\begin{align}
& \asbm_i(z,\gamma) \!=\! \asbm_i(z)\!=\! \frac{q_{-i}}{Q_{\rm m}}(z_i - l_i)-\frac{\eps}{Q_{\rm m}}(z_{-i} -l_{-i} ),
\label{eq:OptimalDrift_Monopolist}
\\
& b_{i,j}^{{\rm sb,m}}(z,\gamma)
\!=\! b_{i,j}^{{\rm sb,m}}(\gamma_i)
\!=\! \Big(\frac{2\Phi_{i,j} }{\gamma_i^{-}}\Big)^\frac14 \!\!\wedge \sigma_{i,j},
~ 
i,j=1,2,
\label{eq:Application_SIATS_OptimalBeta}
\end{align}
with the convention $\frac10=\infty$ and where $-i$ designates the other technology.
In particular, the instantaneous volatility is not impacted by the risk cross-payments $\gamma_3$, so we may restrict to $\Gamma_3=0$ in the contract representation $Y_T^{Z,\Gamma}$. We now solve the regulator problem by using the reduction result of 
\cite{Sannikov2008} and \cite{Cvitanic2018}:
$$
V_{\rm P}^{\rm m}
=
\sup_{Y_0\ge R_0^{\rm m}}
U_{\rm P}^{\rm m},
~\mbox{with}~
U_{\rm P}^{\rm m}
:=
\sup_{Z,\Gamma} 
\E^{\widehat\P} \Big[U_{\rm P}\Big(\!\!-Y^{Z,\Gamma}_T 
                                    \!+\! \int_{0}^{T} \!\!(k^p_1X_t^1 \!+\! k^p_2X_t^2) \dd t
               - \frac{h}{2}\langle X_1\!+\!X_2\rangle_T\Big)
    \Big],
$$
with $\widehat\P$ dynamics for the energy production process $X$ given by
\begin{align}\label{Xhat}
 dX_t^i=\left(a_i^{\rm sb,m}(Z_t)-\delta_i X_t^i\right)\dd t
             +b_{i,1}^{\rm sb,m}(\Gamma^i_t)\dd W_t^{i,1} + b_{i,2}^{\rm sb,m}(\Gamma^i_t)\dd W_t^{i,2}\;, \quad i=1,2.
\end{align}
and dynamics of the additional state $Y$ defined in \eqref{eq:xi1}. 

\subsection{The Results}

We now provide the optimal payment rates which define the regulatory contract offered to the energy producer. Detailed computations are provided in Appendix~\ref{app:oneprod}. Define  
\begin{equation}\label{wi}
w(t) := (w_1(t),w_2(t))^\tp
~\mbox{with}~
w_i(t)=\frac{k_i}{\delta_i}\big(1-e^{-\delta_i(T-t)}\big),
\end{equation} 
and
\begin{align}\label{eq:A1m}
&
% L := \begin{pmatrix} l_1 \\ l_2 \end{pmatrix}, \quad
%  G_{\rm m} := 
%\begin{pmatrix} 
% q_1  &  \eps  \\
% \eps   & q_2  \\
%\end{pmatrix}, \quad
  A^{\rm m}_1 := 
 \frac1\Qm \begin{pmatrix} 
 q_2  & - \eps  \\
 - \eps   & q_1  \\
\end{pmatrix}\!, \,
 \frac{N(t,\gamma)}{\etaP}:= \begin{pmatrix} 
 w_{1}(t) | b^{\rm sb,m}_1(\gamma)|^2 \\ w_{2}(t) |b^{\rm sb,m}_2(\gamma)|^2  
\end{pmatrix}\!, \,
\frac{M(\gamma)}{\etaA \!+\! \etaP} :=
  \begin{pmatrix} 
 |b^{\rm sb,m}_1(\gamma)|^2 & 0  \\
0  & |b^{\rm sb,m}_2(\gamma)|^2 
\end{pmatrix}\!. 
 \end{align}

\begin{Proposition}{\rm (Single firm second-best contract)} \label{prop:oneprodsb}
\begin{enumerate}[\upshape (i)]
\item
The optimal payments are deterministic functions $\zm(t)$ and $\gm(t)$ given as solutions -- if they exist -- of the system of nonlinear equations:
\begin{align*} 
%z(t) & = \Big(M(\gamma(t)) +  A_1 G A_1\Big)^{-1}\Big(A_1 w(t)   - N(t,\gamma(t))\Big), \\
 \zm(t) & = \Big(M(\gm(t)) +  A^{\rm m}_1\Big)^{-1}\Big(A^{\rm m}_1 w(t)  - N(t,\gm(t))\Big), \\
 % z(t) & =  \Big(\mathbb I_2 + G_{\rm m} M(\gamma(t)) \Big)^{-1}\big[ w(t)    - G_{\rm m} N(t,\gamma(t))\big], \\
%\gamma_i(t) & = m_i(t,z_i(t)) \wedge 0, \quad \text{with} \quad m_i(t,z_i) :=  - \etaP (w_i(t) - z_i(t))^2 - \etaA z_i(t)^2 -h.  
\gm_i(t) & = 
m_i(t,\zm_i(t)), \quad \text{with} \quad m_i(t,z_i) :=  - \etaP (w_i(t) - z_i)^2 - \etaA z_i^2 -h.  
\end{align*}
\item
The second-best optimal incentive mechanism is given by 
$\xim = \ximf + \ximv_1 + \ximv_2$ where $\ximv_i$ are as in \eqref{eq:main:xim} and
\begin{align}\label{eq:xim}
\ximf & = R_0^{\rm m} - \int_{0}^{T}\mathcal{H}_A(X_0,z(t),\gamma(t))\dd t.
\end{align}
\end{enumerate}
\end{Proposition}

The existence of a solution for the system of nonlinear equations will be verified numerically. 

\hs

It is possible to provide a slightly more explicit expression for the incentive payments $z$ (see also Appendix~\ref{app:oneprod}). Denoting $Q_i(\gamma_i) := q_{-i} + \Qm (\etaA+\etaP) |b_i(\gamma_i)|^2$  and $\bQ(\gamma) := Q_i(\gamma_i) Q_{-i}(\gamma_{-i}) - \eps^2 > 0$, we can write $\zm_i(t)$ as
\begin{align*}
\zm_i & = \Big( 1 - \etaA \Qm | b_i |^2\frac{Q_{-i}(\gamma_{-i})}{\bQ(\gamma)}\Big) w_i(t) 
- \eps \etaA \frac{\Qm | b_{-i}|^2}{\bQ(\gamma)} w_{-i}(t).
\end{align*}

As this formula makes it clear, unless the firm is risk-neutral or there is no congestion interaction ($\eps=0$), the time-varying prices of each technology depend on each other characteristics. If the firm is risk-neutral ($\etaA=0$) or if there is no congestion cost  (i.e. $\eps=0$), the incentive provisions per technology are decoupled. In each case, the prices per technology basically boil down to the externality values they represent for the regulator. But, apart from these two extreme cases, the optimal incentives for the energy produced and for the energy production volatility reduction are fully coupled. In other words, the price to be paid to the regulated single firm for each new installed renewable energy generation depends on the volatility cost and the cost of reducing volatility of renewable energy technology itself but also on the conventional technology cost structure. 

Moreover, the absolute value of the payment rate for volatility reduction $| \gm_i|$ is an increasing function of the direct cost of volatility $h$. But, even if the regulator does not incur a direct cost of volatility i.e. $h=0$, the payment rates $\gamma_i$ is non-zero.  Indeed, the expression of $\zm_i$ makes it clear that  that $\zm_i<w_i$ as soon as $\Qm>0$ and thus $m_i(t,\zm_i) = - h - \etaA (\zm_i)^2 - \etaP (w_i(t) - \zm_i)^2$ $<0$ even when $h=0$.

\hs

In the setting of Proposition~\ref{prop:oneprodsb}, we can also derive 
the regulator's second-best value function $V_{\rm P}^{{\rm sb,m}} = U_{\rm P}\big(-R_0^{\rm m}+\vsbm(0,X_0^1,X_0^2)\big)$ where
\begin{align}
&\vsbm(t,x_1,x_2) \!=\!\!  \int_{t}^{T} w_0^{\rm m}(s)\dd s +w_1(t) x_1+w_2(t) x_2, 
~\text{with}~ w_0^{\rm m}(t)=  \! \!  \int_t^T \HcP^{\rm m}(\zm(t),\gm(t)) \dd t, \label{eq:vsbm} 
\end{align}
and  $\HcP^{\rm m}(\zm(t),\gm(t))=\sup_{z ,\gamma \in \R^2}\HcP^{\rm m}(t,z,\gamma)$ with
\begin{align}
\HcP^{\rm m}(t,z,\gamma)
&  =
\sum_{i=1,2}  \!\!\asbm_i(z)  w_{i}(t) \!-\! g_i(\asbm(z)) 
\!-\! \phi_i\big(b_i^{{\rm sb,m}}(\gamma)\big)    
 \!+\! \frac12 |b_i^{{\rm sb,m}}(\gamma) |^2 m_i(t,z_i)) .
 \end{align}
For the sake of comparison, we report here the explicit solution of the corresponding business-as-usual problem corresponding to the non-incentivized firm case, and the first-best solution where the regulator chooses the actions of the firm while preserving its reservation utility. 

\begin{Proposition}{\rm (Single firm business-as-usual case)}
\label{prop:Application_SIATS_NoContract}
The firm's optimal investment rates  without contract are given by
\begin{align}
& a_j^{{\rm bu,m}}(t) =
\frac{q_i}{Q_{\rm m}}(w_j^{{\rm A}}(t)-l_j) - \frac{\eps}{Q_{\rm m}} (w_i^{{\rm A}}(t)-l_i), \quad
b_{j,l}^{{\rm bu,m}}(t)
 = \Big(\frac{2\Phi_{j,l}}{\eta_{\rm A} w_j^{A}(t)^2}\Big)^\frac14 \wedge \sigma_{j,l}, \label{eq:OptimalDriftMonopolist}  \\
&\text{with}  \quad w_j^{{\rm A}}(t) = \frac{p}{\delta_j}\big(1-e^{-\delta_j(T-t)}\big), \quad j=1,2, \quad j\neq i.
 \end{align}
\end{Proposition}

Without contract, the firm's value function is given by 
\begin{align}
&V_{\rm A}(0)=U_A\left(w_1^{{\rm A}}(0)X_0^1 + w_2^{{\rm A}}(0)X_0^2 + \int_{0}^{T} w_0^{{\rm bu,m}}(t)\dd t\right), \quad \text{where} \nonumber \\
&w^{{\rm bu,m}}_0(t) =\sum_{j=1}^{2}\Big\{ w_j^{{\rm A}}(t) a_j^{{\rm bu,m}}(t)
-\frac12 \eta_{\rm A} w_j^{{\rm A}}(t)^2  |b_j^{{\rm bu,m}}(t)|^2 
-g_j( a_j^{{\rm bu,m}}(t))-\sum_{l=1}^{2}\phi_{j,l}( b_{j,l}^{{\rm bu,m}}(t))\Big\}.
\end{align}

Apart from the depreciation rate, we see as expected that the optimal investment strategy of the energy producer without incentives is driven by the price $p$ of electricity. Basically, the firm increases capacity in technology $j$ only if the price $p$ is larger than the per unit value of investment $l_j$. And, when the firm is risk-averse, it undertakes some volatility reduction efforts.

\hs

Moreover, by using \cite{Lin24} Proposition 3.1 (2), we may also deduce from Proposition~\ref{prop:oneprodsb} the characterization of the corresponding first-best problem, which is obtained in the limiting risk-neutral Agent setting $\eta_{\rm A}=0$. 

\begin{Proposition}{\rm (Single firm first-best optimum)} \label{prop:FB_Monopolist}
If the firm is risk-neutral, then the optimal payment rates for the capacities and their quadratic variations reduce to $z_i(t) = w_i(t)$ and $\gamma_i(t) = - h$ for $i=1,2$.
The optimal prices for the contract's variable part  to the risk-neutral firm simplify to
\begin{align}
\pim_i(t) = k_i - p, \quad \rimi(t) = - h.
\end{align}
\end{Proposition}

Note, that the first-best optimal prices are the regulator's marginal values of energy investments  and of intermittency for each energy source. Besides, the regulated firm only receives  the externality values of the technologies ($k_1$ and $k_2$) for each deviation from the baseline and does no longer benefit from the market price $p$  since the price is substracted from the contract. Indeed, the only guarantees the regulator has to provide to the firm is its reservation utility.  Thus, the regulator only compensates the firm for its costs. Regarding the volatility, the situation is even  simpler: the regulated single energy producer is taxed at a level price $h$, inducing him to reduce the variability of the production until the marginal abatement cost equals the volatility cost $h$. Note that the result above applies even if there is some congestion cost between the two technologies.

\section{Solution of the regulator facing two competing producers} \label{sec:twoprod}

\subsection{The Contract}
\label{sec:Hi}

In the case where the development of the two technologies described in Section~\ref{sec:oneprod} is under the control of two firms investing each in one technology, we consider the lower bound $\underline{V}_{\rm P}^{{\rm sb,c}}$ introduced in \eqref{eq:underline-VF_Principal_TIATS}. The two agents are now offered contracts $\xi^i=Y^i_T$ where $Y^i$ has dynamics of the form
\be\label{Yi}
\dd Y^i_t
&=&
Z^{i,.}_t \cdot \dd X_t+\frac12\sum_{k=1,2}(\Gamma^{i,k}_t+\eta_i|Z^{i,k}_t|^2)|b^k_t|^2\,\dd t
-\Hc_i(X_t,Z^i_t,\Gamma^i_t) \dd t,~i=1,2,
\ee
where the maximization in the agents Hamiltonians can be separated as in the previous section: 
\begin{align}
\Hc_i(x,z,\gamma) & =   \Hc^0_i(x,z) + \overline\Hc_i(z) +  \langle\Hc\rangle_i(\gamma), \quad
 \text{with} \quad
 \Hc^0_i(x,z) = (p-\delta_i z_{i,i})x_i - \delta_j x_j z_{i,j},
 \label{Hi0}\\
 \overline\Hc_i(z) &
 = a_{j} z_{i,j}+\max_{a_i} \big\{a_i z_{i,i} - g_i(a)\big\}, \quad
 \langle\Hc\rangle_i(\gamma) 
 := 
\frac12 |b_j|^2 \gamma_{i,j} + \sup_{b_i} \Big\{\frac12 | b_i|^2 \gamma_{i,i} - \phi_i(b_i)\Big\}.
 \end{align}
As standard, all Nash equilibria can be characterized through the corresponding Nash equilibria at the level of the Hamiltonians which we now analyze for the optimal choice of drift and diffusion:
\begin{enumerate}
\item[(i)] Because there is no interaction between the actions of the firms $i$ on the volatility of the dynamics of production capacity of firm $j$, the best-response functions $b_i(\gamma)$ are the same as in the single firm case. Observing that the term $\langle X_1+X_2\rangle$ in the regulator's objective function \eqref{JPc} does not involve $\Gamma^{1,2},\Gamma^{2,1}$, it follows that the regulator's optimal choice for these variables is $\gamma^{i,j}_t =0$ for all $t\in[0,T]$. 
\item[(ii)] Due to the coupling term $\eps a_1a_2$ in the cost function $g_i(a)$, we immediately verify that there is a unique Nash equilibrium $\ac_i(z)$ for the payment functions $(a_1 z_{1,1} - g_1(a), a_2 z_{2,2} - g_2(a))$, characterized by the first order conditions $z_i-q_ia_i-l_i-\frac12\eps a_j=0$, $i=1,2$ and the notation $j=2\1_{\{i=1\}}+\1_{\{i=2\}}$. This linear $2\times 2$ system of linear equations provides
\begin{align*}
\ac_i(z) = \frac{q_{j}}{\Qc}(z_{i,i} - l_i) - \frac12 \frac{\eps}{\Qc}(z_{j,j} - l_{j}), \quad \Qc := q_i q_{j} - \frac14 \eps^2.
\end{align*}
\end{enumerate}
Hence, the unique Nash equilibrium at the level of the Hamiltonian induces the payments $\xi^i=Y^i_T$ where the processes $Y^i$ are defined by the dynamics \eqref{Yi} with equilibrium Hamiltonian defined by \eqref{Hi0} and
\begin{equation}
\overline\Hc_i(z) 
=
\bar h_i(\ac(z),z):=   a^{\rm c}_i(z) z_{i,i} + a^{\rm c}_j(z) z_{i,j} - g_i\big(a^{\rm c}_i(z)\big),
~\mbox{and}~
\langle\Hc\rangle_i(\gamma) =  \frac12 | b_i(\gamma_i)|^2 \gamma   - \phi_i(b_i(\gamma_i)),
\end{equation}
where $\gamma_i:=\gamma_{i,i}$ for notational simplicity.

Now, similar to the single agent setting, we introduce a subset of incentive mechanisms defined through arbitrary processes $Z := (Z^{1,1},Z^{2,2},Z^{1,2},Z^{2,1})^\top$ and $\Gamma := (\Gamma^1,\Gamma^2)^\top$: 
\begin{align}\label{eq:xi2}
Y_T^{i,Z, \Gamma} = &
Y_0^i 
  + \int_{0}^{T} \!\!\!Z^{i,i}_t \dd X^i_t + Z^{i,j}_t \dd X^j_t
                        -\Hc_i(X_t,Z_t,\Gamma_t) \dd t \nonumber \\
                       & \hspace{4mm} + \frac12 \int_{0}^{T}  \big\{\big[\etai (Z^{i,i}_t)^2 
                                              \!+\! \Gamma^i_t\big] | b_i(\Gamma^i_t)|^2  \!+\! \etai (Z^{i,j}_t)^2 | b_j(\Gamma^j_t)|^2 \big\} \dd t.
\\ 
= &
Y_0^i 
  + \int_{0}^{T} \Big(g_i(\ac(Z_t)) + \phi_i(b_i(\Gamma^i_t)) - p X^i_t 
  + \frac12 \etai \big[  (Z^{i,i}_t)^2 | b_i(\Gamma^i_t)|^2 +  (Z^{i,j}_t)^2 | b_j(\Gamma^j_t)|^2\big]  \Big)\dd t 
\nonumber
\\
  & + \int_0^T Z^{i,i}_t  b_i(\Gamma^i_t) \cdot \dd W^i_t + Z^{i,j}_t  b_j(\Gamma^j_t) \cdot \dd W^j_t. \nonumber
\end{align}

This expression makes clear that despite the fact that the best-responses functions $b_i(\gamma_i)$ of each firm on the volatility reduction are the same as in the single firm case, the optimal incentive payment rates may be different because the regulator has also to compensate firm $i$ for the volatility cost induced by the payment rate $Z^{i,j}$. 

By following the same line of argument as in the Principal-(single)-Agent problem, it follows that  
$$
V_{\rm P}^{{\rm sb,c}}
\;\ge\;
\underline{V}_{\rm P}^{{\rm sb,c}}
\;:=\;
\sup_{Y_0^n\ge U_n(R_0^{{\rm c},i})}
\;\sup_{Z,\Gamma} J_{\rm P}\big(Y_T^{i,Z, \Gamma}, \P^{\hat\nu(Z,\Gamma)}\big),
$$
where $\hat\nu(Z,\Gamma)=\big(a^{\rm c}(Z),b^{\rm c}(\Gamma)\big)$ with $a^{\rm c}(Z)$ the above defined Nash equilibrium in point~(ii) and $b_i^\star(\Gamma)=b_i^{{\rm sb,m}}(\Gamma)$, $i=1,2$, as defined in \eqref{eq:Application_SIATS_OptimalBeta}.

In contrast with the single agent setting of the Principal-Agent problem, we do not know whether the equality between $V_{\rm P}^{{\rm sb,c}}$ and $\underline{V}_{\rm P}^{{\rm sb,c}}$ holds. This is indeed an open problem which requires further developments of multi-dimensional second order backward SDEs. However, we shall see later that $\underline{V}_{\rm P}^{{\rm sb,c}}\ge V_{\rm P}^{{\rm sb,m}}$, which implies the remarkable result announced in the introduction section and stated in Theorem~\ref{thm:comparison} that $V_{\rm P}^{{\rm sb,c}}\ge V_{\rm P}^{{\rm sb,m}}$.

We next comment on the form of the contractual incentive mechanism~\eqref{eq:xi2}. First, the same interpretations done in the case of a single firm in Section~\ref{ssec:contract1} applies to the payment rates $Z^{i,i}$. But, in the case of two interacting firms, the regulator provides to firm~$i$ a payment $Z^{i,j} \dd X^{j}$ even though firm~$i$ has no control over the dynamics of the capacity of the technology~$j$. These cross-payments allow the regulator to influence the Nash equilibrium of the interacting firms.

 Second, regarding the payment rates for volatility reduction, there are no differences compared to the case of one single firm. Indeed, because there are no co-variation between $X^1$ and $X^2$ and there are no interactions in the cost functions $\phi_i$, the expressions of the best-responses of each firm $i$ to payment rates $\gamma^i$ are expected to be the same as in the single firm case. Nevertheless, at equilibrium, the reduction effort on volatilities should be different because they depend also on the payment rates~$z$.

 Finally, it is not clear if the regulator can achieve a higher value in the case of the regulation of one single firm controling the two technologies or in the case of two firms in interaction. Indeed, intuitively, the regulation of two interacting firms forming a Nash equilibrium seems to lead to a lower value compared to a single firm case. But, in the case of two interacting firms, the regulator has more control variables than in the single firm case. Indeed, in the single firm case, she has four control variables: two payment rates for the capacities and two for their quadratic variations. In the present case of two interacting firms, the regulator has more control variables: four payment rates for the capacities and the same number of payment rates for the quadratic variations. It turns out that the regulator can take advantage of this larger number of control variables to achieve a higher value in the latter case.

\subsection{The Results}\label{ssec:results2}

Define the $2\times 2$ and $2\times 4$ matrices together of the vector in $\R^4$:
\begin{equation}\label{eq:A1c}
G_{\rm c} := \begin{pmatrix}
q_1 & \frac{\eps}2 \\
\frac{\eps}2 & q_2
\end{pmatrix}, ~
A^{\rm c}_1 := 
 \big(
 G^{-1} | 0_{2\times 2} 
 \big),
 ~
 \frac{N_{\rm c}(t,\gamma)}{\etaP} 
 := 
w_{1}(t) | b_1^{\rm sb,c}(\gamma)|^2 
\begin{pmatrix} 1\\ 0\\ 0 \\ 1 \end{pmatrix}
+w_{2}(t) |b_2^{\rm sb,c}(\gamma)|^2 \begin{pmatrix} 0\\ 1\\ 1 \\ 0 \end{pmatrix},
 \end{equation}
and the $4\times 4$ matrix
\begin{align}\label{eq:Mc}
&\!\!M_{\rm c}(\gamma) \!\!:= \!\!
  \begin{pmatrix} 
 (\eta_1 \!\!+  \!\!\etaP) |b_1^{\rm sb,c}(\gamma)|^2 & \hspace{-5mm} 0 & \hspace{-5mm}  0 & \hspace{-5mm}  \etaP | b_1^{\rm sb,c}(\gamma)|^2 \\
0  & \hspace{-5mm}  (\eta_2  \!\!+  \!\!\etaP) |b_2^{\rm sb,c}(\gamma)|^2 & \hspace{-5mm}  \etaP |b_2^{\rm sb,c}(\gamma)|^2  & \hspace{-5mm}  0 \\
0 & \etaP |b_2^{\rm sb,c}(\gamma)|^2 & \hspace{-3mm}  (\eta_1+\etaP) |b_2^{\rm sb,c}(\gamma)|^2 & \hspace{-5mm}  0 \\
\etaP | b_1^{\rm sb,c}(\gamma)|^2 & \hspace{-5mm}  0& \hspace{-5mm}  0 & \hspace{-7mm}   (\eta_2+\etaP)|b_1^{\rm sb,c}(\gamma)|^2 \\
\end{pmatrix}. 
 \end{align}
where the functions $w_i$, $i=1,2$ have been defined in Proposition~\ref{prop:oneprodsb}.
 
\begin{Proposition}{\rm (Two firms second-best contract)} \label{prop:twoprodsb}
\begin{enumerate}[\upshape (i)]
\item
The optimal payments are deterministic functions, $\zc(t) \in \R^4$ and $\gc(t) \in \R^2$,  given as solutions -- if they exist -- to the  system of nonlinear equations:
\begin{align}\label{eq:zgsb-c}
\zc(t) &  = \big(M_{\rm c}(\gc(t)) +  (A_1^{\rm c})^\tp G A_1^{\rm c}\big)^{-1} \big\{ (A_1^{\rm c})^\tp \big(w(t)   + \big[G G_{\rm c}^{-1} -\mathbb I_2\big]L\big)  - N_{\rm c}(t,\gc(t)) \big\}, \\
\gc_i(t)  & = 
 -  \etai \zc_{i,i}(t)^2 -\etaj  \zc_{j,i}(t)^2 - \etaP (\zc_{i,i}(t) + \zc_{j,i}(t) - w_i(t))^2 - h =: m_i^{\rm c}(t,\zc(t)),
\end{align}
inducing optimal investment efforts $\ac_i(\zc(t))$ and identical volatility reduction efforts $b^{\rm c}_i(\gamma_i) = b^{\rm sb,m}_{i}(\gamma_i)$.

\item
The second-best optimal contract for producer $i$ is given
 by $\xic_i = \xicf_i + \xicv_{i,i} + \xicv_{i,j}$ where
\begin{align}
&\xicf_i =   R_0^{{\rm c},i} - \int_{0}^{T}  \Hc_i(X_0,z(t),\gamma(t))\dd t, \\
&\xicv_{i,j} =  
\int_{0}^{T} \pic_{i,j}(t) (X_t^j - X_0^j) \dd t 
- \frac12 \int_{0}^{T} \ricij(t) \dd \langle X^j\rangle_t 
+ (X_T^j - X_0^j) z_{i,j}(T), \\
 &
\pic_{i,i}(t) = -\dot{z}^{\rm c}_{i,i}(t) - p + \delta_i \zc_{i,i}(t), \quad
\pic_{i,j}(t)  = -\dot{z}^{\rm c}_{i,j}(t)   + \delta_j \zc_{i,j}(t),  \label{eq:SBpricesDrift} \\
& \ricii(t) =  
 h + \etaj \zc_{j,i}(t)^2 + \eta_{\rm P} \Big(\zc_{i,i}(t) + \zc_{j,i}(t) - w_i(t)\Big)^2,
\quad
\ricij(t)= - \etai \zc_{i,j}(t)^2.
\end{align}
\end{enumerate}
\end{Proposition}

\hs

In the presence of incentives, the Principal's value function is given by 
$$V_P^{{\rm sb,c}} = U_{\rm P}\Big(- R_0^{{\rm c},1} -R_0^{{\rm c},2} + \vsbc(0,X_0^1,X_0^2)\Big)$$ with
\begin{align*}
& v^{\rm c}(t,x_1,x_2) = w_0^{\rm c}(t) + w_1(t) x_1 + w_2(t) x_2,  \quad
  w_0^{\rm c}(t) = \int_t^T \HcP^{\rm c}(\zc(t),\gc(t)) \dd t, \\
 & \HcP^{\rm c}(\zc(t),\gc(t)) = \sup_{z \in \R^4, \gamma \in \R^2} \Big\{
 \sum_{i=1,2}  \ac_i(z)  w_i(t) - g_i(\ac(z)) - \phi_i(b_i)    
 + \frac12 |b_i |^2 m_i^{\rm c}(t,z) \Big\}.
 \end{align*}

\hs

 It is clear from the Proposition~\ref{prop:twoprodsb} above that when there is no congestion cost (i.e. $\eps = 0$) and when $\eta_{\rm A} = \eta_1 = \eta_2$, the firms' best-response correspond to the best-response of the single firm for each technology. Indeed, in that case, firms only respond to their own direct payment rates $z_{i,i}$. In that case, it is clear that the certainty equivalent $\vsbm$ and $\vsbc$ are equal. 

%\subsection{Benchmarks}\label{ssec:2firm-bench}
 
 \hs
 
As in the single firm case, we report here the explicit solutions of the corresponding business-as-usual problem and the case of risk-neutral firms.

 \begin{Proposition}{\rm (Two firms business-as-usual)}
\label{prop:twofirmsbau}
In the absence of incentives, the optimal investment rate $a^{{\rm bu,c}}_i$ and volatility reduction effort ${b}^{{\rm bu,c}}_i$ of firm $i$ are given by
\begin{align}
&a^{{\rm bu,c}}_i(t) = \frac{q_j}{Q_{\rm c}} (w_i^{{\rm A}}(t) - l_i) - \frac{\eps}{2Q_{\rm c}} (w_j^{{\rm A}}(t)-l_j), \quad
{b}^{{\rm bu,c}}_{i,k}(t) = \Big(\frac{2\Phi_{i,k}}{\eta_i w_i^{{\rm A}}(t)^2}\Big)^\frac14 \wedge \sigma_{i,k}, \quad k=1,2,  \\
&\text{with} \quad   
w_i^{{\rm A}}(t) = \frac{p}{\delta_i}\big(1-e^{-\delta_i(T-t)}\big).
\end{align}
\end{Proposition}

\hs

In the absence of incentives, the value function of the firm~$i$ is given by
\begin{align}
& V_i(0) = U_i\Big(w_i^{{\rm A}}(0) X_0^i + \int_{0}^{T} w_{i,0}^{{\rm bu,c}}(s) \dd s \Big) \nonumber \\
&\text{with}  \quad w_{i,0}^{{\rm bu,c}}(t) = w_i^{{\rm A}}(t)  a^{{\rm bu,c}}_i(t) - 
\frac12 \eta_i | b^{{\rm bu,c}}_i(t) |^2 w_i^{{\rm A}}(t)^2 - C_i( a^{{\rm bu,c}}(t),b^{{\rm bu,c}}(t)). 
\end{align}

\hs

Moreover, even if it is still an open question whether the case of risk-neutrality of interacting Agents provides the first-best optimum in the case with volatility control, we provide this benchmark solution in the next proposition. We define 
\begin{align*}
& \zeta_i :=\frac{q_j}{\Qc}\left(1-\frac{\eps^2}{2 \Qc}\right),  \quad
 %   F_{i,i} :=\frac{q_j}{\Qc} + \frac{\mathcal{E}}{\zeta_j}\frac{\eps}{2\Qc}, \quad
    F_{i,i} := \frac{q_j}{\Qc} + \frac{\eps^4}{4 q_i \Qc} \frac{1}{2\Qc -  \eps^2}, \quad
    F_{i,j} := \frac{\eps}{2\Qc} + \frac{\eps^3}{2\Qc} \frac{1}{2\Qc -  \eps^2}, \\
%&    F^0_{i} := K_i - K_j \frac{\mathcal{E}}{\zeta_j}, \quad
&    F^0_{i} := K_i - K_j \frac{\eps^3}{2q_i} \frac{1}{2\Qc - \eps^2}, \quad
 K_i : = \frac{\eps}{2\Qc} \Big[l_j  \left( 1 + \frac{\eps^2}{2\Qc}\right) - l_i q_j \frac{\eps}{\Qc}\Big], \quad
   \zeta := \zeta_i \zeta_j - \frac{\eps^6}{16 \Qc^4}.
\end{align*}

\begin{Proposition}{\rm (Risk-neutral firms)}
\begin{enumerate}[\upshape (i)]
\item If both producers are risk-neutral, the payments rates for capacity of firm $i$ are given by
\begin{align}
&z_{i,i}(t) = w_i^{P}(t)-z_{j,i}(t), \quad
 z_{i,j}(t) = z_{i,j}(t)=
w_j(t) \Big(1 - \frac{\zeta_i}{\zeta } F_{j,j} \Big)
+  \frac{ \zeta_i }{ \zeta  } \Big(w_i(t) F_{j,i} - F^0_{j}\Big).
\end{align}
The payments for quadratic variations are constant given by $\gamma_{i} = - h$.
\item The second-best optimal prices for quadratic variation for both producers reduce to $\ricii = - h$ and $\ricij = 0$.
\end{enumerate}
\end{Proposition}

In the regulation of interactive firms, we note that even if both firms are risk-neutral, their drift payment rates are coupled. This is in contrast to the case of the regulation of the single energy producer (see  Proposition~\ref{prop:FB_Monopolist}). 
 
\section{Numerical Findings} \label{sec:Application_Numerics}

In this section, we analyze numerically our theoretical findings in the single and two energy producer setting with and without volatility control. Our model conjugates several features which makes the analysis of their relative effects complex. Thus, Section~\ref{sec:num-drift} limits the analysis to the case of incentive provision on capacity only and for a single firm. Section~\ref{sec:num-vol} shows the benefit from an incentive for volatility reduction, still in a single firm setting. Section~\ref{sec:num-inter} provides the effect of the interaction of two firms compared to a single firm when there is incentives only for capacity and not for volatility. Finally, Section~\ref{sec:num-vol-inter} provides a view of the joint effect of volatility reduction incentive in a interacting firms context. Sections~\ref{sec:num-prod} and \ref{sec:num-contract} provides respectively comparisons of the level of production and the values of the payments.

We consider a reference case for which all modeling parameters can be found in Tables~\ref{tab:FixedParameterValue_Agents}, \ref{tab:FixedParameterCost_Agents} and \ref{tab:FixedParameter_Regulator}.

%\subsection{The Reference Case}
\begin{table}[htp]
\centering
\begin{tabular}{ccccc}
$X_0$  & $dt$  & $\delta$ &  $p$ &  $\eta_1=\eta_2=\eta_{\rm A}$ \\
\midrule
$[5000,1000]$  & $\frac{1}{52}$& $[0,0]$ & $100$ &  $0.001$ \\
\end{tabular}%
\caption{Parameter values for the state and value function of the energy producer(s).}
\label{tab:FixedParameterValue_Agents}%
\end{table}
\begin{table}[htp]
\centering
\begin{tabular}{ccccccc}
$l$ &  $q$  & $\eps$ & $\Phi_{1,\cdot }$&$\Phi_{2,\cdot }$ & $\sigma_{1,\cdot }$ & $\sigma_{2,\cdot }$    \\
\midrule
 $[100,200]$  & $[1,2]$ & $0.25$  & $[2000^4,6000^4]$ & $[5000^4,6000^4]$ & $[250,100]$ & $[500,100]$    \\
\end{tabular}%
\caption{Parameter values for the cost function of the energy producer(s).}
\label{tab:FixedParameterCost_Agents}%
\end{table}
\begin{table}[htp]
\centering
\begin{tabular}{cccc}
$T$ & $\eta_{\rm P}$ & $k$ & $h$\\
\midrule
$10$ &$0.001$ & $[-1,800]$ & $50^4$  \\
\end{tabular}%
\caption{Parameter values of the regulator.}
\label{tab:FixedParameter_Regulator}%
\end{table}
Note that the parameters $p,l,k$ are in \euro/MWh and $h,q,\eps,\Phi$ in \euro/MWh$^2$ so that we  multiply in the numerics with 24 hours and 7 days in a week as $\dd t$ identifies weekly steps.

Throughout this section, we use the shortcuts DC and DVC to differentiate situations where the incentive mechanism involves only control of the drift (DC) and when it involves both drift and volatility control (DVC).

Finally, all the results presented in this section consider $R^{\rm m}_0 = R^{\rm c}_1 = R^{\rm c}_2 =0$.

\subsection{Absence of volatility control and interaction} \label{sec:num-drift} % Note, that h\neq 0 here!

%\begin{figure}[htp] %htp
%\centering
%\includegraphics[width=0.7\textwidth]{Monopolist_NoContract_DC_ReferenceCase.eps}\\[2ex]
%\includegraphics[width=0.7\textwidth]{Monopolist_SBContract_DC_ReferenceCase.eps}
%\caption{Monopolist: BU and SB without Volatility control}
%\label{fig:Monopolist_NoV_BU-SB}
%\end{figure}

%\begin{figure}[htp] %htp
%\centering
%\includegraphics[width=0.7\textwidth]{Monopolist_SBContract_DC_ReferenceCase.eps}\\[2ex]
%\includegraphics[width=0.7\textwidth]{Monopolist_SBContract_DVC_ReferenceCase.eps}
%\caption{Monopolist: SB with and without Volatility control}
%\label{fig:Monopolist_SB_NoV-VC}
%\end{figure}

\begin{figure}[htb]
\centering
\begin{subfigure}[b]{0.45\textwidth}
	\centering
 \caption{Investment effort}
	\includegraphics[width=0.9\textwidth]{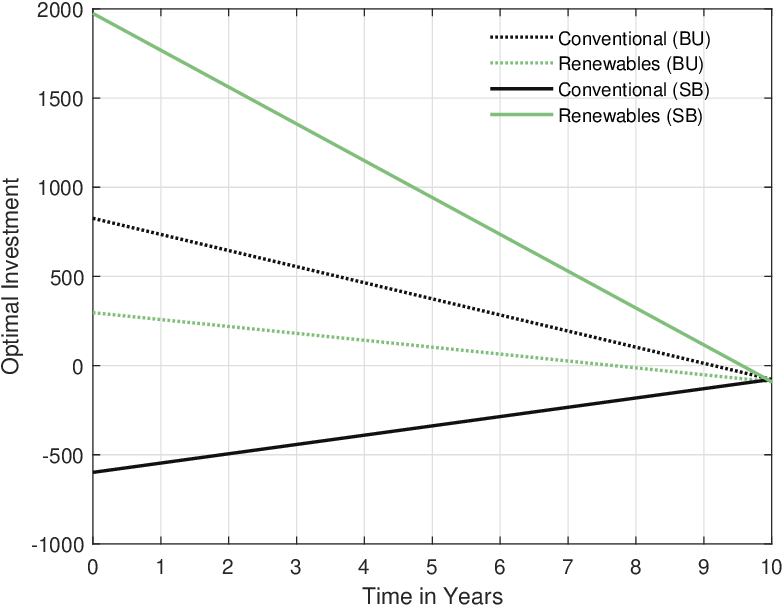}
\end{subfigure}\hfill
\begin{subfigure}[b]{0.45\textwidth}
\centering
\caption{State process}
 \includegraphics[width=0.9\textwidth]{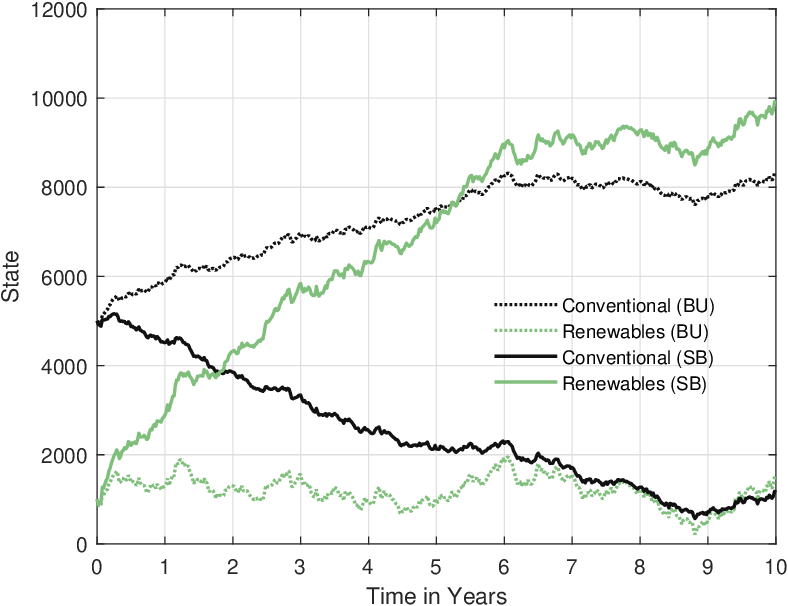}
\end{subfigure}%
\caption{A comparison of business-as-usual (BU) and second-best (SB) implications for the single producer setting without volatility control, in which the constant volatilities, $\sigma_{1,\cdot }$ and $\sigma_{2,\cdot }$, are taken from \ref{tab:FixedParameterCost_Agents}. (a): Investment efforts $a^{{\rm sb,m}}$ in energy production of the single firm. (b): Energy production in the single producer setting.}
\label{fig:M_DC}
\end{figure}

Figure~\ref{fig:M_DC}~(a) illustrates the optimal investments  rates  for the single firm as a function of time.
We observe in black  the optimal investment rate $a_1^{{\rm sb, m}}$ in conventional technology and in green,  the investment rate $a_2^{{\rm sb,m}}$ in renewable technology. The dotted lines represent the behaviors under business-as-usual (BU), while the solid lines refer to the second-best~(SB) responses.  In the business-as-usual case, the investment rate in conventional technology is positive and larger than the one in renewable. This situation reflects the fact that in the absence of regulation, a firm prefers to invest in a cheaper yet polluting technology compared to a more expensive one yet clean technology. Moreover, these investments  rates decrease linearly with time at different paces (large investment effort at the beginning of the period, small at the end). This phenomenon arises since 
\begin{align}
    \dot{ a}_j^{{\rm bu,m}}=-\frac{p(q_i-\eps)}{Q_{\rm m}}<0\;, \quad \quad \text{for } j=1,2, \quad j\neq i,
\end{align}
 where we recall that $\Qm = q_1 q_2 - \eps^2$. Since $\eps<q_1<q_2$, the slope for conventional energy is steeper. 

In contrast,  in the presence of incentive mechanism, the investment rate in conventional technology is negative over the whole period of time, meaning that the firm reduces the total capacity of emissive production ($k_1<0$). Moreover, it is linearly increasing since
\begin{align}
    \dot{ a }_1^{{\rm sb,m}}= \frac{q_2}{Q_{\rm m}}\dot{z}_1 - \frac{\eps}{Q_{\rm m}}\dot{z}_2 >0.
\end{align}
This translates that the disinvestment effort is larger at the beginning of the period than at the end. On the other side, the investment rate in renewable technology is drastically larger compared to the BU because $k_2>0$ and $\dot{ {a}}_2^{{\rm sb,m}}<0$. Note, that since $\delta_1=\delta_2=0$, \ak{the marginal optimal drift payments} $\dot{z}_1$ and $\dot{z}_2$ become constant. Moreover, we observe that the terminal controls, $\asbm_j(T)= {a}_j^{{\rm bu,m}}(T)$, coincide at terminal time since \mbox{$z_j(T)=w^{{\rm A}}_j(T)=0$} for $j=1,2$. 

Figure~\ref{fig:M_DC}~(b) illustrates the dynamics of the capacities for renewable and conventional  technologies  with investment rates control and incentive for capacities only.  The green and black trajectories depict renewable and conventional production capacity, respectively.  The dotted lines again characterize the evolution in the business-as-usual setting and the solid lines specify the evolution under the second-best contract. 

As a consequence of the different investment rates shown in Figure~\ref{fig:M_DC}~(a), we observe that under business-as-usual, the production capacity in conventional emissive technology increases with time whereas it decreases in the presence of incentives. Moreover, incentives has an accelarating effect in the development of renewable capacity as it is drastically larger than in the BU.

Finally, we observe higher fluctuations for renewable than for conventional in both situations~(business-as-usual and second-best) since $\sigma_{2,l}\geq\sigma_{1,l}$ for $l=1,2$. In particular, the renewables' volatility, $\sigma_{2,1}$, is  twice as high as for conventional production, $\sigma_{1,1}$.

\subsection{Volatility control and absence of interaction}  \label{sec:num-vol}

In this section, we focus on the setting with a single energy producer who can perform investment both in capacity and in volatility reduction.

\begin{figure}[htb]
\centering
\begin{subfigure}[b]{0.45\textwidth}
	\centering
	\caption{Optimal volatility response under BU}
	\includegraphics[width=0.9\textwidth]{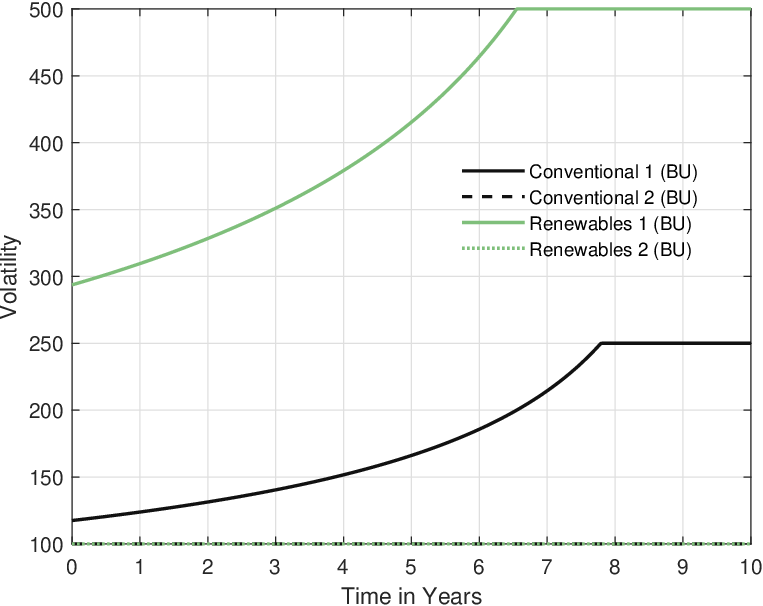}
\end{subfigure}\hfill
\begin{subfigure}[b]{0.45\textwidth}
\centering
\caption{Optimal volatility response under SB}
\includegraphics[width=0.9\textwidth]{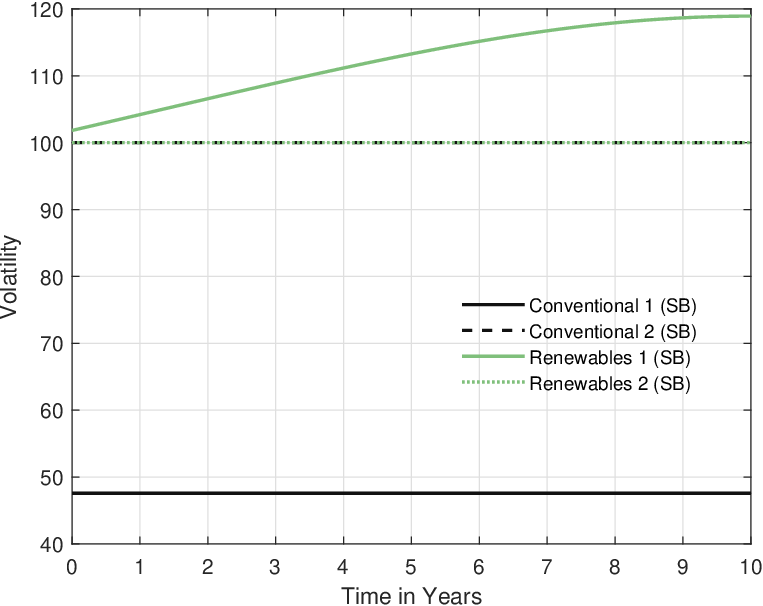}
\end{subfigure}%
\caption{The single producer setting with volatility control. (a): Firm's optimal volatility control $b^{{\rm bu,m}}$ in the business-as-usual case. (b): Firm's optimal volatility control $b^{{\rm sb,m}}$ under the second-best contract.}
\label{fig:M_SB_Vola}
\end{figure}
\begin{figure}[htb]
\centering
\begin{subfigure}[b]{0.45\textwidth}
	\centering
	\caption{Investment effort}
	\includegraphics[width=0.9\textwidth]{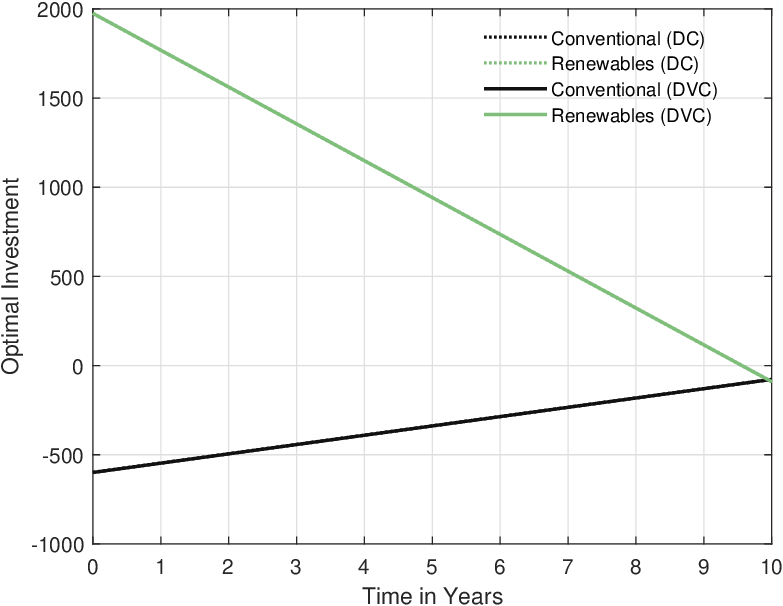}
\end{subfigure}\hfill
\begin{subfigure}[b]{0.45\textwidth}
\centering
\caption{State process}
\includegraphics[width=0.9\textwidth]{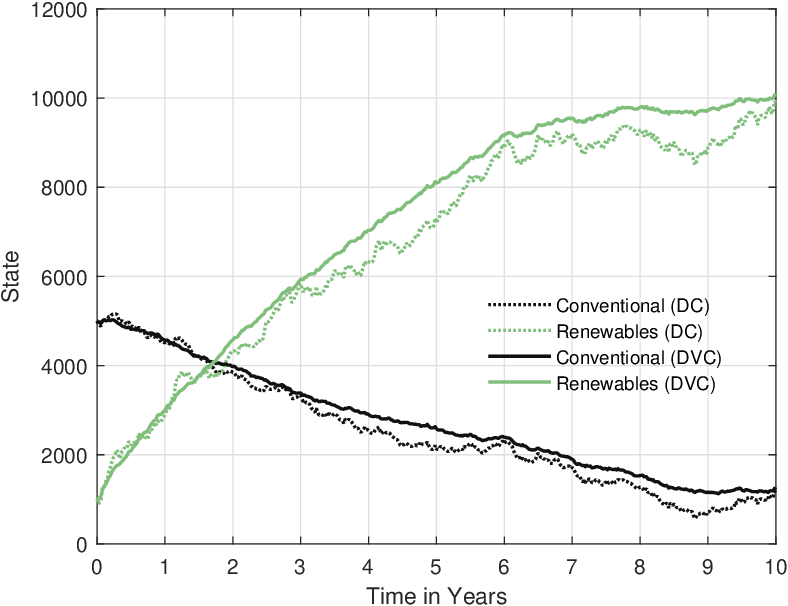}
\end{subfigure}%
\caption{The single producer setting under the second-best contract with volatility control (DVC) and without volatility control (DC, i.e. with constant uncontrolled volatilities $\sigma_{1,\cdot }$ and $\sigma_{2,\cdot }$  given by Table~\ref{tab:FixedParameterCost_Agents}). (a): Firm's optimal investments $\asbm_j$ in energy production. (b): Energy production in the single producer setting.}
\label{fig:M_SB}
\end{figure}

Figure~\ref{fig:M_SB_Vola} illustrates the single firm's volatility control in the business-as-usual scenario and under the second-best contract over the whole time horizon. The solid and dotted green lines refer to the controlled volatility of renewables. The solid and dashed black lines refer to the controlled volatility of conventional energy production. 

We make the following observations. First, even the absence of incentives and of direct cost of volatility, the firm undertakes some volatility reduction effort (Figure~\label{fig:M_SB}~(a)) because it is a risk-averse firm. These efforts are different as a function of the source of shocks: the firm performs no reduction effort for the second Brownian $W^{.,2}$ because we choose a large cost of reduction. And, the effort decreases with time because $b^{{\rm bu,m}}_{j,l}(t) \sim (T-t)^{-1/4}$ and it stops as soon as $b^{{\rm bu,m}}_{j,l}(t) = \sigma_{j,l}$.  Second, volatility reduction incentives provide the desired effect on some sources of shocks: no effort is still performed on the volatilities of $W^{.,2}$ for both technologies but significant reduction is obtained for the other source of shocks. This translates in the smoothing of the capacities trajectories of both technologies as shown in Figure~\ref{fig:M_SB}~(b). Third, the introduction of incentive to reduce volatility does not significantly modify the investment rates in both technology. Figure~\ref{fig:M_SB}~(a) shows investment rates of the single firm in the presence of incentives with and without volatility incentive reduction. They are undistinguishable. This result should reassure regulators: the introduction of new policies to reduce intermittency should not have a significant impact on the investment pace in renewable technology and on the disinvestment in emissive technology.

\subsection{Interaction in the absence of volatility control}  \label{sec:num-inter}

\begin{figure}[htb]
\centering
\begin{subfigure}[b]{0.45\textwidth}
	\centering
	\caption{Investment effort under BU}
	\includegraphics[width=0.9\textwidth]{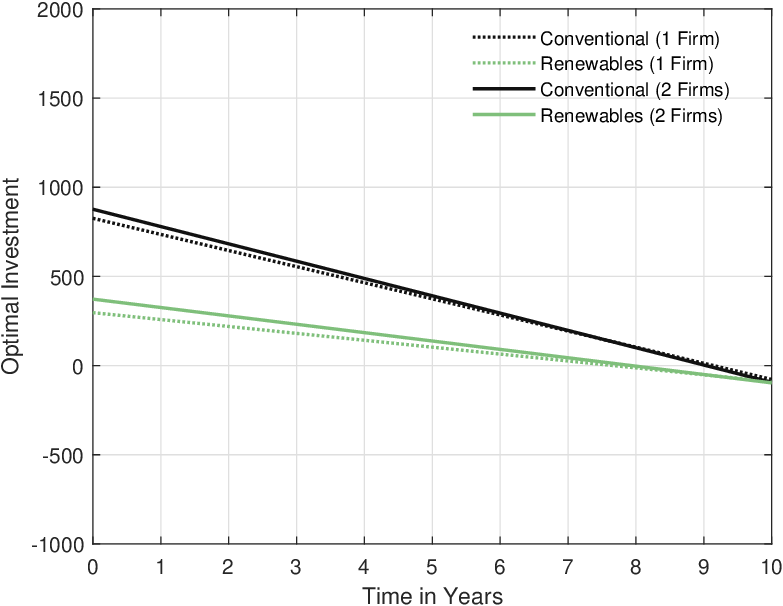}
\end{subfigure}\hfill
\begin{subfigure}[b]{0.45\textwidth}
\centering
\caption{State process under BU}
\includegraphics[width=0.9\textwidth]{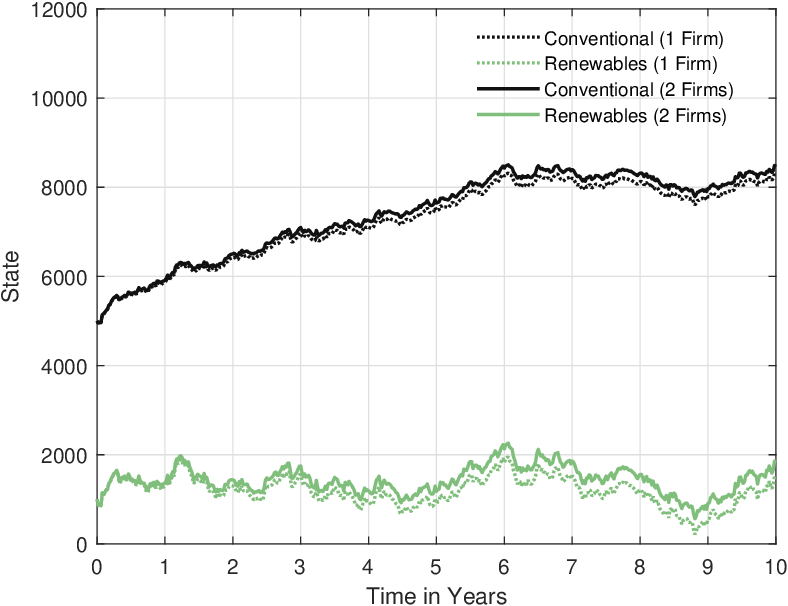}
\end{subfigure}%
\caption{A comparison of the market structures,  one (M) and two firms (C), in the business-as-usual case without volatility control. (a): Investment efforts $a^{{\rm bu},\cdot}$. (b): Energy production.}
\label{fig:Comparison_DriftState_BU_DC}
\end{figure}

\begin{figure}[htb]
\centering
\begin{subfigure}[b]{0.45\textwidth}
	\centering
	\caption{Investment effort under SB}
	\includegraphics[width=0.9\textwidth]{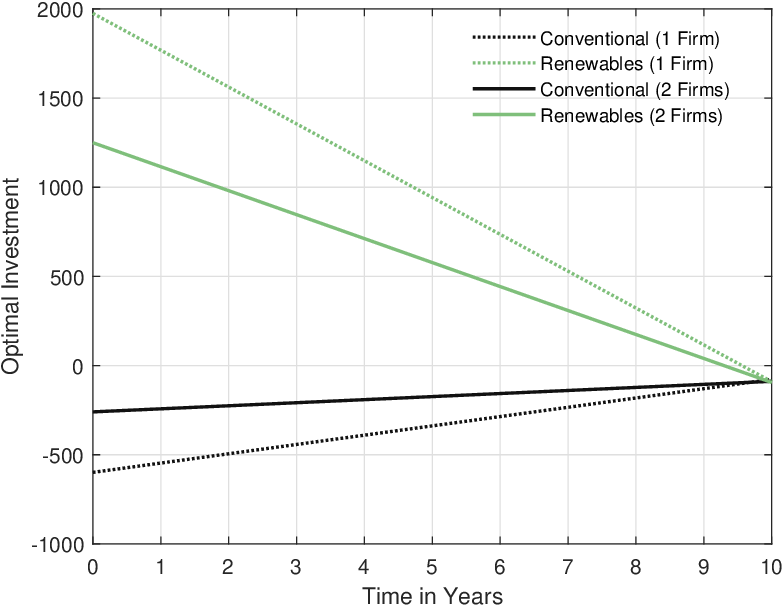}
\end{subfigure}\hfill
\begin{subfigure}[b]{0.45\textwidth}
\centering
\caption{State process under SB}
\includegraphics[width=0.9\textwidth]{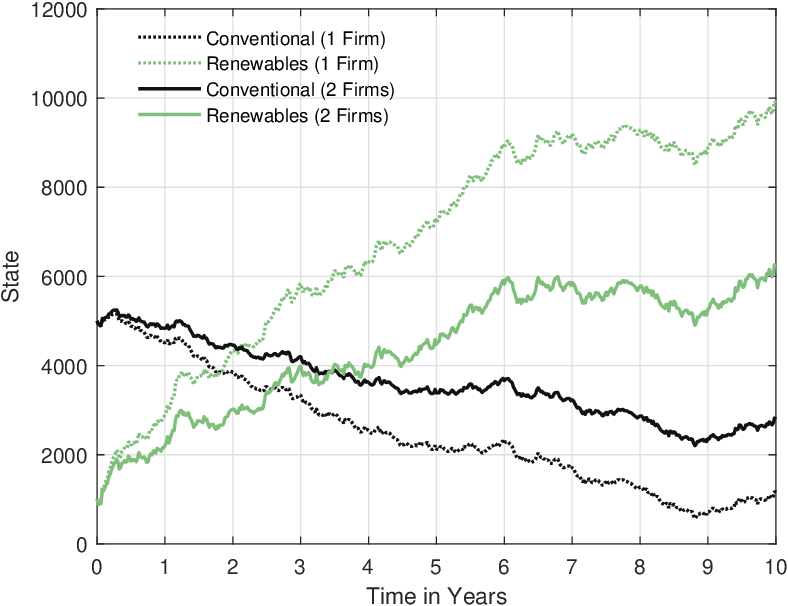}
\end{subfigure}%
\caption{A comparison of the market structures,  one (M) and two firms (C), under a second-best contract without volatility control. (a): Investment efforts $a^{{\rm sb},\cdot}$. (b): Energy production.}
\label{fig:Comparison_DriftState_SB_DC}
\end{figure}

Figures~\ref{fig:Comparison_DriftState_BU_DC} and \ref{fig:Comparison_DriftState_SB_DC}  illustrate the optimal investments and the resulting state processes without volatility control in the business-as-usual case (\ref{fig:Comparison_DriftState_BU_DC}) and under the second-best contract (\ref{fig:Comparison_DriftState_SB_DC}), respectively.
As before, green lines refer to renewable energy and black lines are assigned to conventional energy. The dotted lines indicate a market structure with one energy producer whereas the solid line indicates the interactive setting with two firms.


First, we observe that the market structure has little effect on the investment decisions in the business as usual (Figure~\ref{fig:Comparison_DriftState_BU_DC}~(a) and \ref{fig:Comparison_DriftState_BU_DC}~(b)) whereas it has a significant impact when it comes to incentives provision (Figure~\ref{fig:Comparison_DriftState_SB_DC}~(a) and \ref{fig:Comparison_DriftState_SB_DC}~(b)). The regulator achieves a larger renewable capacity in the case of a single firm (10~GW) compared to the case of interacting firms case (6~GW); and a larger disinvestment in emissive technology with a single firm  compared to two interacting firms (1~GW of capacity in the former compared to 3~GW for the latter). 

\subsection{Volatility control and interaction}
\label{sec:num-vol-inter}

\begin{figure}[htb]
\centering
\begin{subfigure}[b]{0.45\textwidth}
	\centering
	\caption{Optimal volatility response under BU}
	\includegraphics[width=0.9\textwidth]{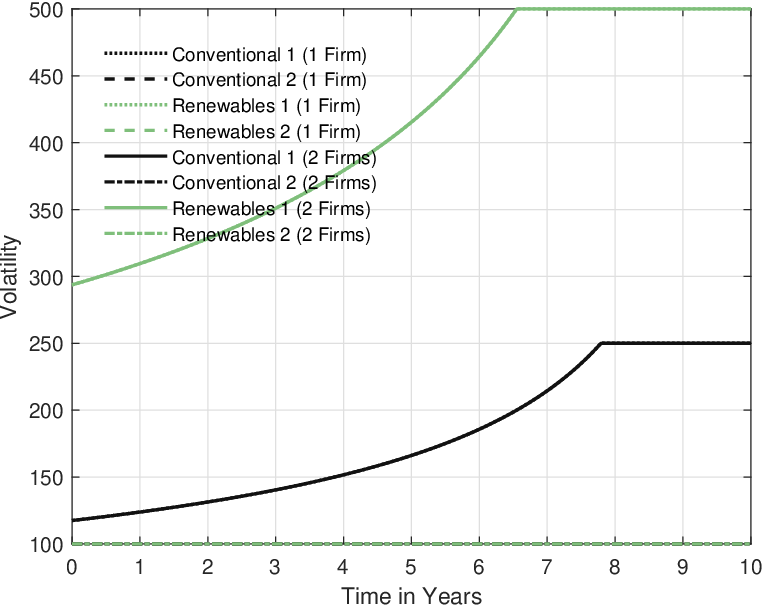}
\end{subfigure}\hfill
\begin{subfigure}[b]{0.45\textwidth}
\centering
\caption{Optimal volatility response under SB}
\includegraphics[width=0.9\textwidth]{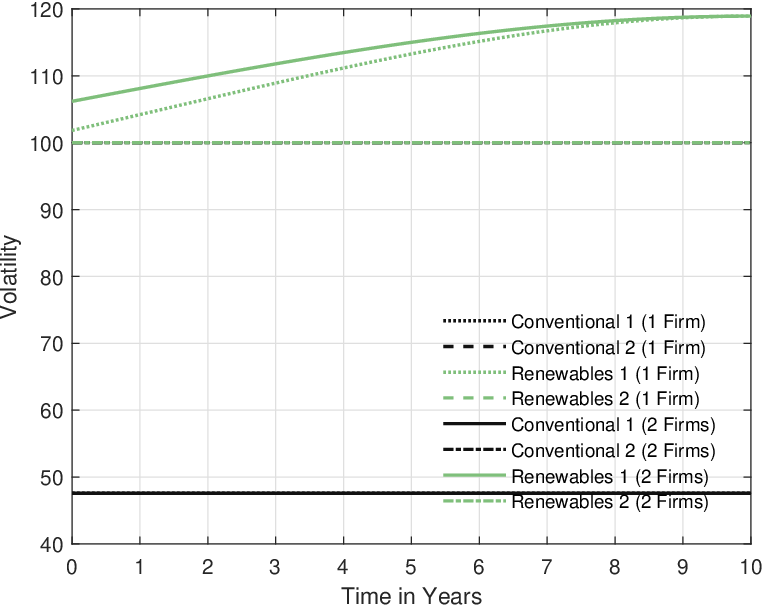}
\end{subfigure}%
\caption{A comparison of volatility controls for both market structures, one (M) and two firms~(C). Subfigure~(a): Optimal volatility in the business-as-usual case. Subfigure~(b): Optimal volatility under the second-best contract.}
\label{fig:C_DVC_Vola}
\end{figure}

In Figure~\ref{fig:C_DVC_Vola}, we consider the same setting as in Figure~\ref{fig:M_SB_Vola} in dashed and dotted lines and supplement the interactive setting in solid and dash-dotted lines.  As in Figure~\ref{fig:M_SB_Vola}, there is no volatility reduction effort for the second souce of shock $W^{\cdot,2}$ in any cases. For the remaining volatilities, we observe that in the absence of a contract, the optimal volatility responses coincide (see Figure~\ref{fig:C_DVC_Vola}~(a)), since $b^{{\rm bu,m}}_{j,l}= b^{{\rm bu,c}}_{j,l}$ for $j,l=1,2$. However, under a second-best contract, the volatility efforts differ between the market structures (see Figure~\ref{fig:C_DVC_Vola}~(b)) as the volatility payments $\gamma_j \neq \gamma^j_{j}$ for $j=1,2$ do. For both technologies, we find that $b^{{\rm sb,m}}_{j,l}\leq b^{{\rm sb,c}}_{j,l}$ since $\gamma_j\leq \gamma^j_{j}$ for $j,l=1,2$ for nearly the whole time horizon. For the last month only, this behavior changes to $\gamma^j_{j}<\gamma_j$ reaching $\gamma^j_{j}(T)<\gamma_j(T)=-h$ at terminal time for $j=1,2$.

\begin{figure}[htb]
\centering
\begin{subfigure}[b]{0.45\textwidth}
\centering
\caption{State process under BU}
\includegraphics[width=0.9\textwidth]{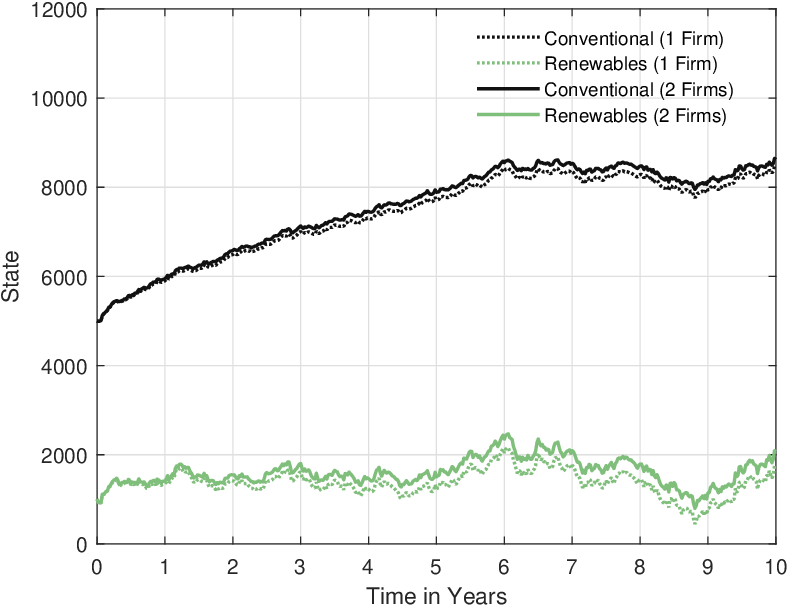}
\end{subfigure}\hfill
\begin{subfigure}[b]{0.45\textwidth}
\centering
\caption{State process under SB}
\includegraphics[width=0.9\textwidth]{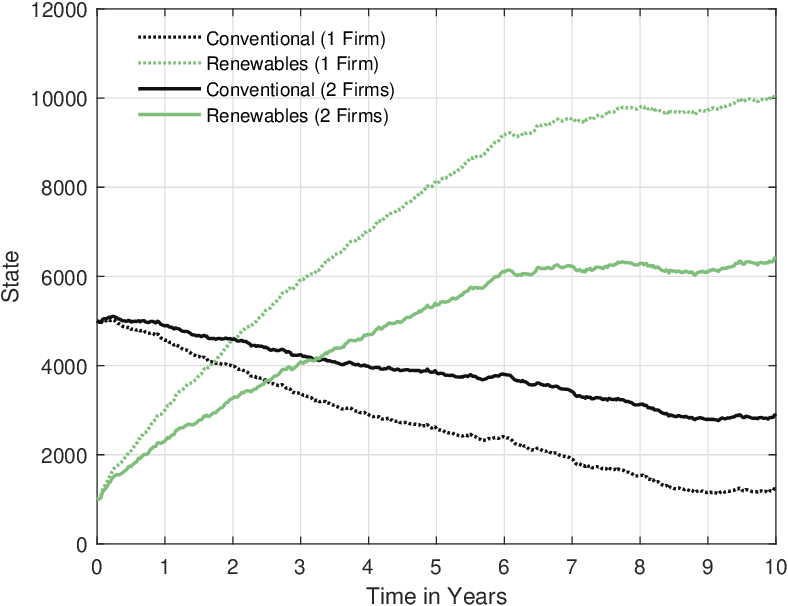}
\end{subfigure}%
\caption{A comparison of the energy production with volatility control for both market structures, one (M) and two firms (C). Subfigure~(a): Energy production in the business-as-usual case. Subfigure~(b): Energy production under the second-best contract.}
\label{fig:C_DVC}
\end{figure}

From Figure~\ref{fig:M_SB}, we already now that the effect of volatility efforts on the optimal investment level is small compared to its absence. Hence, we refer to Figure~\ref{fig:Comparison_DriftState_BU_DC}~(a) and Figure~\ref{fig:Comparison_DriftState_SB_DC}~(a) for a comparison of optimal investments between the market structures in the case with and without a second-best contract, respectively.

Figure~\ref{fig:C_DVC} shows the dynamics of the production capacities of the two technologies in the case of two interacting firms and volatility control and incentives reduction. Even in the absence of a contract (see Figure~\ref{fig:C_DVC}~(a)), the states under volatility control are  slightly smoother than without volatility control (see Figure~\ref{fig:Comparison_DriftState_BU_DC}~(b)) due to the fact that the producers are slightly encouraged to stabilize the production even in the absence of a contract. 

In Figure~\ref{fig:C_DVC}~(b), the joint effect manifests more explicitly. We observe that the trajectories from Figure~\ref{fig:Comparison_DriftState_SB_DC}~(b) become a smoothed version~(as in Figure~\ref{fig:M_SB}~(b) for the single producer case). In fact, the dotted lines of Figure~\ref{fig:C_DVC}~(b) coincide with the solid lines from Figure~\ref{fig:M_SB}~(b). We observe, as in  Section~\ref{sec:num-inter}, that interaction leads to less pronounced production development. Moreover, as in Section~\ref{sec:num-vol}, the ability of controlling the state's variability leads to higher production  on the long-term than under drift control only.

\subsection{Production} \label{sec:num-prod}

\begin{figure}[htb]
\centering
\begin{subfigure}[b]{0.45\textwidth}
	\centering
	\caption{Total production}
	\includegraphics[width=0.9\textwidth]{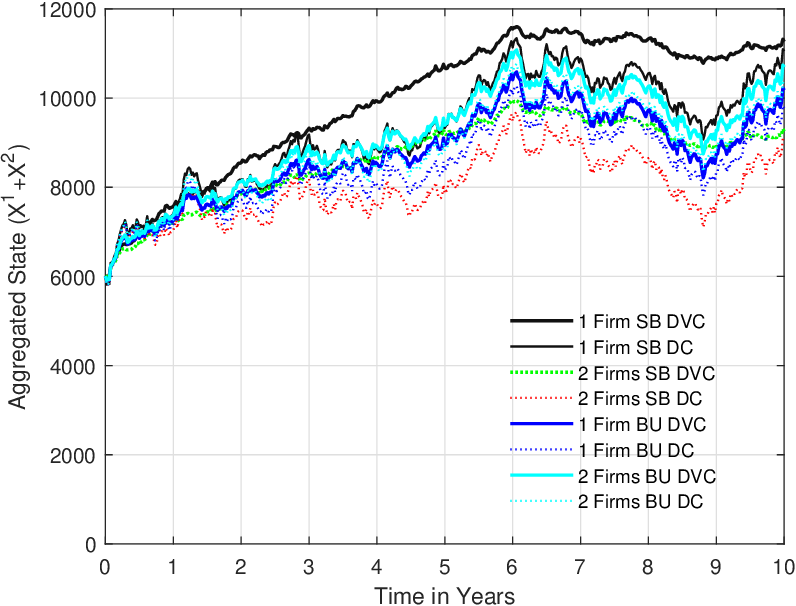} 
\end{subfigure}\hfill
% \begin{subfigure}[b]{0.3\textwidth}
% \centering
% \caption{Renewable production capacity}
% \includegraphics[width=0.9\textwidth]{RenewableState_Comparison_ReferenceCase_Paper_20231229.eps}
% \end{subfigure}
% \hfill
\begin{subfigure}[b]{0.45\textwidth}
	\centering
	\caption{Share of renewables' production}
	\includegraphics[width=0.87\textwidth]{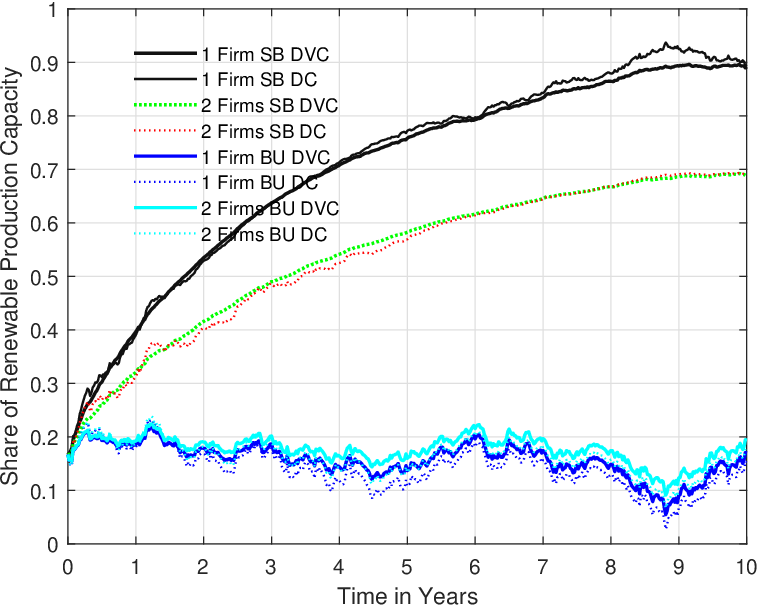} 
\end{subfigure}%
\caption{Development of energy production resulting from both market structures (M and C) with and without volatility control (DVC and DC) in the absence and presence of the second-best contract (BU and~SB). Subfigure (a): Total energy production. 
%Subfigure (b): Renewable energy production capacity. 
Subfigure (b): Share of renewables' production. }
\label{fig:TotalEnergyProduction}
\end{figure}

Figure~\ref{fig:TotalEnergyProduction} summarizes the development of the total energy production and the share of renewables'   for all scenarios resulting from different market structures (M and C), from business-as-usual and second-best contracts (BU and SB), and the absence and presence of volatility control (DC and DVC). In Figure~\ref{fig:TotalEnergyProduction}~(a), all scenarios are endowed with a total initial production of 6~GW and rise over the  time horizon of ten years.

On the long term, the  market structure with a single producer regulated by the second-best contract outperforms all other scenarios with more than 11~GW under investment rate and volatility control and approximately 10~GW under investment rate  control only. The interactive setting with  investment rate control only leads under the second-best contract to the lowest capacity with around 9~GW at the end of the contracting period. Interaction under a second-best scenario with  investment rate  and volatility control is even in line with all business-as-usual scenarios on the long-term, however, with a drastic difference: the share of renewable energy production capacity in this regulated interactive setting is around 70\% (in contrast to its business-as-usual scenario with approximately 17\% share of renewables; see Figure~\ref{fig:TotalEnergyProduction}~(b)).
In a market structure with one producer, the share is even higher at around 90\% at the end of the contracting period (in contrast to its business-as-usual scenario with 14\% share of renewables).
Hence, this second-best contract is able to capture the long-term goals of the Paris Climate Agreement. 
Without regulatory action the share of renewable production  remains below 20\%  on the long-term independent of the scenario.

\subsection{Contracts} \label{sec:num-contract}

We denote $\xi_M^{DC}$, $\xi_{C,1}^{DC}+\xi_{C,2}^{DC}$, $\xi_M^{DVC}$ and $\xi_{C,1}^{DVC}+\xi_{C,2}^{DVC}$, the expected contract value paid by the regulator in the one (M) and two firms (C) setting under drift control only (DC) and with drift and volatility control (DVC). In Table~\ref{tab:Contracts}, we observe the following ranking of contracts: $\xi_M^{DC}>\xi_{C,1}^{DC}+\xi_{C,2}^{DC}>\xi_M^{DVC}>\xi_{C,1}^{DVC}+\xi_{C,2}^{DVC}$.
Hence, we find that the regulator pays more, whenever the volatility cannot be controlled since there are missing tax earnings in the uncontrolled settings.
The scenario, in which the volatility can be controlled, is less expensive since the regulator taxes the variability of the production.

Moreover,  the sum of both interactive contracts is less expensive than the contract of a single producer. This effect appears on the one hand through
existing cross payments or charges, $z_{12}$ and $z_{21}$, respectively.
On the other hand,
interaction between the producers diminishes the peculiarity of investment effects for which regulatory subsidies are paid or charges are taken, respectively. 
Hence, the single producer setting with more pronounced investment efforts are more costly for the regulator.

\begin{table}[htb]
\begin{center}
\begin{tabular}{cccc}
$\xi_M^{DC}$ &	$\xi_M^{DVC}$	& $\xi_{C,1}^{DC}+\xi_{C,2}^{DC}$	& $\xi_{C,1}^{DVC}+\xi_{C,2}^{DVC}$ \\	
 \midrule
 % 4.24e+14 &
 % 1.02e+14 &
 % 3.77e+14 &
 % 9.87e+13
 1.96e+14&	1.05e+14&	1.74e+14&	1.03e+14
\end{tabular}
\caption{Total contract value paid by the regulator in the one (M) and two firm (C) setting under drift control only (DC) and with drift and volatility control (DVC).}
\label{tab:Contracts}
\end{center}
\end{table}

\section{Conclusion}\label{sec:Conclusion}

We considered general Principal-Multi-Agent incentive problems with hidden drift and volatility actions and a lump-sum payment at the end of the contracting period. The contribution is threefold: first, we show how investments in renewable energy production can be increased while simultaneously ensuring a stable energy production in a single firm setup and in an interacting setting with two agents. Second, we provide a case where the regulation of two firms in interaction can achieve a higher social value compared to the regulation of a single firm handling the two technologies. The numerical study highlights the impact of the contract design on the investments in (non-)renewable energy production. This will lead to significant benefits for the regulator as well as to great adjustments in average investment and responsiveness of the power producer.

Furthermore, the development of applications of optimal incentive mechanism in continuous-time with volatility control are scarce. This paper offers an example outside the fields of financial economics where the design of incentive to reduce volatility of the system is a natural and desirable objective. Moreover, the recent work of \cite{Chiusolo24} offers a new alternative to the approach of optimal incentive with volatility control that can potentialy reduce the supplementary layer of computational complexity involved with the control of volatility, allowing for more applications in the future.

%
% Appendices
%

%\newpage
\appendix

\section{Detailed computations for Section~\ref{sec:model}} \label{app:model}

\subsection{Regulator facing a single producer}\label{app:oneprod}

%\myblue{Moreover, we note that the reduction effort on volatility and its associated cost function are the same as in \cite{Aid2022}. Indeed, in that paper, the authors takes as volatility function $\sigma_i(u_i)$ and cost function $c_i(u_i)$ for a given Brownian $i$:
%\begin{align*}
%\sigma_i(u_i) := \sqrt{u_i} \sigma_i, \quad 
%c_i(u_i) := \frac{\sigma_i^2}{\lambda_i}\Big(\frac{1}{u_i} - 1\Big), \quad
%u_i \in (0, 1].
%\end{align*}
%Taking $b_i := \sqrt{u_i} \sigma_i$ and $\lambda_i := \sigma_i^4/\Phi_i$ ensures the correspondance. We use this feature to state the optimal volatility reduction effort and refer to \cite{Aid2022} for the technical details required to fully express the conditions under which the control is legitimate, in particular the fact that the reduction effort should be bounded away from zero.}  

%In this appendix devoted to the second-best optimal contract of the regulator facing a single producer, we drop all subscripts to reduce the notational burden. 

We recall that the producer's best-response to incentive payment rates $z$ and $\gamma$ are
\begin{align*}
a_i(z) = \frac{q_{-i}}{Q_{\rm m}}(z_i- l_i) - \frac{\eps}{Q_{\rm m}}(z_{-i}- l_{-i}), \quad
b_{i,j}(\gamma) = \Big(\frac{2\Phi_{i,j}}{\gamma_i^-}\Big)^\frac14 \wedge \sigma_{i,j}, \quad
i,j=1,2, \quad
\Qm =q_1q_2-\eps^2,
\end{align*}
inducing the dynamics
\begin{align*}
\dd X^i_t & = \big(a_i(Z_t) -\delta_i X_t\big) \dd t 
+ b_{i,1}(\Gamma_t^i) \dd W^{i,1}_t 
+ b_{i,2}(\Gamma_t^i) \dd W^{i,2}_t.
\end{align*}
The dynamics of the contract $Y$ satisfies
\begin{align*}
\dd Y_t = \sum_{i=1}^2 \Big(g_i(a(Z_t)) + \phi_{i}(b_{i}(\Gamma^i_t)) + \frac12 \etaA (Z^i_t)^2 |b_i(\Gamma^i_t)|^2 - p X^i_t\Big) \dd t
+ Z^i_t\big(b_{i,1}(\Gamma_t^i) \dd W^{i,1}_t 
+ b_{i,2}(\Gamma_t^i) \dd W^{i,2}_t\big).
\end{align*}
The HJB of the value function of the regulator $V(t,x,y)$ then is
\begin{align*}
-\partial_t V = \sup_{z,\gamma}\Big\{ &
\sum_{i=1,2} \big(g_i(a(z)) + \phi_i(b_i(\gamma_i)) + \frac12 \etaA z_i^2 |b_i(\gamma_i)|^2 - px_i \big) V_y 
+ (a_i(z) - \delta_i x_i) V_{x_i}\\
&+ \frac12 z_i^2 |b_i(\gamma_i)|^2 V_{yy} 
 + \frac12 |b_i(\gamma_i)|^2 V_{x_ix_i} 
 + z_i |b_i(\gamma_i)|^2 V_{yx_i} \\
& - \etaP \big[(k_i-p)x_i - \frac12 h |b_i(\gamma_i)|^2 \big] V
\Big\}, \quad
V(T,x,y) = U_{\rm P}(-y).
\end{align*}
We follow the standard approach by a solution of this equation of the form $V(t,x,y) = U_{\rm P}(v(t,x) - y)$. Direct calculation reduces to the following equation in $v$:
\begin{align*}
-\partial_t v = \sup_{z,\gamma}\Big\{ &
\sum_{i=1,2} -\big(g_i(a(z)) + \phi_i(b_i(\gamma_i)) + \frac12 \etaA z_i^2 |b_i(\gamma_i)|^2 - px_i \big)
+ (a_i(z) - \delta_i x_i) v_{x_i}  \\
& - \frac12 \etaP z_i^2 |b_i(\gamma_i)|^2  
 + \frac12 |b_i(\gamma_i)|^2 (v_{x_ix_i} - \etaP v_{x_i}^2) 
 + \etaP z_i |b_i(\gamma_i)|^2 v_{x_i}\\
& + (k_i-p)x_i - \frac12 h |b_i(\gamma_i)|^2 
\Big\}, \quad
v(T,x) = 0.
\end{align*}
Thus, dropping implicit dependencies on $z$ and $\gamma$ in $a$ and $b$, the HJB rewrites
\begin{align*}
-\partial_t v =& \sum_{i=1,2} (k_i-\delta_i v_{x_i}) x_i   + \sup_{z,\gamma} \HcP(z,\gamma) \\
\text{with} &  \quad   \HcP(z,\gamma)  := 
\sum_{i=1,2}  a_i  v_{x_i} - g_i(a) - \phi_i(b_i)    
 + \frac12 |b_i |^2 m_i(z,v_{x_i}, v_{x_ix_i}), \\
\text{and} & \quad m_i(z,v_i, v_{ii}) := v_{ii} -  \etaA  z_i^2 - \etaP (z_i - v_i)^2  - h.
\end{align*}

Optimisation w.r.t the volatility reduction can be performed for each $b_{i,j}$ independently for fixed $z$. Besides, optimization can be done first w.r.t. $b_{i,j}$ taking care of the constraint that $b_{i,j} \in (0, \sigma_{i,j}]$, and then deduce the incentive payment rates $\gamma_i$. Making the change of variable $u_{i,j} := b_{i,j}^2$ and noting that
$ \frac{2\Phi_{i,j}}{u_{i,j}^2} = \gamma_i^-$, first-order condition provides
\begin{align*}
 v_{x_ix_i} - \etaP (v_{x_i} -z_i)^2 - h - \etaA z_i^2 + \gamma_i^- = 0.
 \end{align*}
Hence, to ensure $b_{i,j} \leq \sigma_{i,j}$, the optimal incentive payment rates is 
\begin{align*}
\gamma_i = m_i(z,v_{x_i},v_{x_ix_i}) \wedge 0. 
 \end{align*}

\hs 
Now, optimization over $z =(z_{1}, z_{2})^\intercal$ resolves in finding the maximiser $z$ of the function
\begin{align}\label{eq:G}
& H_{\rm p}(z,\gamma)  :=  Dv \cdot a(z) - L \cdot a(z) - \frac12 a(z) \cdot G a(z) - \frac12 z \cdot M(\gamma) z -  N(\gamma) \cdot z, \quad  \text{with} \nonumber \\ 
 & a(z) = A^{\rm m}_1 (z - L), \quad 
 L := (l_1, l_2)^\tp, \quad
 G := \begin{pmatrix}
q_1 &  \eps \\
 \eps & q_2
\end{pmatrix} = (A_1^{\rm m})^{-1}, \quad
 \end{align}
and where $A_1^{\rm m}$, $L$, $M(\gamma)$ and $N(\gamma)$ have been defined in~\eqref{eq:A1m}. Besides, we have $Dv \cdot   A_1 z$  $= (A_1^{\rm m} z)^\intercal Dv$ $ = A_1^{\rm m} Dv \cdot z$, because $A_1^{\rm m}$ is symmetric. Besides,
\begin{align*}
a(z) \cdot G a(z) & 
%=  A_1 z \cdot G A_1 z + A_1 z \cdot G A_0 + A_0 \cdot G A_1 z + A_0 \cdot G A_0 \\
 = z \cdot A_1^{\rm m}  z -  2 A_1 L \cdot z + A_1^{\rm m} L  \cdot L
\end{align*}
Hence, first-order condition w.r.t. $z$ provides
\begin{align*}
A_1^{\rm m} (Dv - L) -  A_1^{\rm m} z +  A_1^{\rm m} L - M(\gamma) z - N(\gamma) =0, \\
\big(M(\gamma) +  A_1^{\rm m}\big) z = A_1^{\rm m} Dv - N(\gamma). 
\end{align*}
Assuming $M(\gamma) + A_1^{\rm m}$ to be invertible, the optimal incentives rates are given by the system of equations:
\begin{align*}
z = \Big(M(\gamma) +  A_1^{\rm m}\Big)^{-1}\Big(A_1^{\rm m} Dv  - N(\gamma)\Big), \quad
\gamma_i = m_i(z,v_{x_i},v_{x_ix_i}) \wedge 0. 
\end{align*}

Now, similar to \cite{Aid2022} making the educated guess that $v(t,x_1,x_2) = w_0(t) + w_1(t) x_1 + w_2(t) x_2$, we get the incentive payment rates are deterministic functions of time $z_i(t)$ and $\gamma_i(t)$ solutions (if they exist) of the following system
\begin{align} 
%z_i(t) & =  \Big( 1 - \etaA \frac{Q_{-i}(\gamma_{-i}) \Qm | b_i |^2}{\bQ(\gamma)}\Big) v_{x_i} - \eps \etaA \frac{\Qm | b_{-i}|^2}{\bQ(\gamma)} v_{x_{-i}}, \label{eq:zgsb-z} \\
z(t) & = \Big(M(\gamma) +  A_1^{\rm m}\Big)^{-1}\Big(A_1^{\rm m} w(t)  - N(\gamma)\Big), \label{eq:zgsb-z} \\
\gamma_i(t) & = 
m_i(t,z_i(t)), \quad \text{with} \quad m_i(t,z_i) :=  -h  - \etaA z_i(t)^2 - \etaP (w_i(t) - z_i(t))^2, \label{eq:zgsb-g}
\end{align}
where $w(t) := (w_1(t) \, w_2(t))$ are determined by the ODEs
$$
- \dot w_i  = k_i - \delta_i w_i, ~~ w_i(T) = 0,~~i=1,2,
~~\mbox{and}~~
 -\dot w_0(t) = \HcP(z(t),\gamma(t)) \dd t,~~w_0(T) = 0,
$$
where $z$, $\gamma$ and  $m$ are given by~\eqref{eq:zgsb-z}-\eqref{eq:zgsb-g}. Finally, by a standard verification argument we may justify that the current solution $V(t,x,y)=U_{\rm P}(v(t,x)-y)$ with the above affine map $v$ is indeed the required value function. 

\hs
 
\noindent We now rewrite the optimal contract in the rebate form given by Proposition~\ref{prop:oneprodsb}~(ii). It holds that 
\begin{align*}
Y_T & = Y_0 + \sum_{i=1}^2 \int_0^T z_i(t) \dd X^i_t + \frac12\big(\gamma_i(t) + \etaA z_i^2(t)) \dd \langle X^i\rangle_t - 
\HcA^i(X^i_t,z(t),\gamma_i(t)) \dd t, \\
\HcA^i(x_i,z,\gamma_i) & := (p-\delta_i z_i) x_i + \underbrace{z_i a_i(z) - g_i(z)}_{\overline\Hc_i(z)} + \underbrace{\frac12 |b_i(\gamma_i)|^2 \gamma_i - \phi_i(b_i(\gamma_i))}_{\langle\Hc_i\rangle(\gamma_i)}.
\end{align*}
By integration by part using the fact that $z_i(T)=0$, we have
\begin{align*}
\int_0^T z_i(t) \dd X^i_t - (p - \delta_i z_i(t)) X^i_i \dd t & = 
\underbrace{-z_i(0) X^i_0  - \int_0^T \dot z_i(t) X^i_0 \dd t}_{=0}  
 + \int_0^T \big[ \underbrace{\delta_i z_i(t) - p -\dot z_i(t)}_{=: \pi_i^{\rm m}(t)} \big] (X^i_t - X^i_0)\dd t  \\
& - \int_0^T \underbrace{(p - \delta_i z_i(t)) X^i_0}_{= \bar \Hc_i(X^i_0,z_i(t))} \dd t.
\end{align*}
Moreover, $\rimi(t) := - \gamma_i(t) - \etaA z_i^2(t) =  h + \etaP(w_i(t) + z_i(t))^2$. Finally, 
\begin{align*}
\int_0^T \big[\overline\Hc_i(z(t)) + \langle\Hc_i\rangle(\gamma_i(t)) \big] \dd t
\end{align*}
is independent of the trajectory of $X$ and thus, enters in the constant payment part of the contract.

\hs

\noindent{\bf Remark}: In this case of the regulation of a single firm, it is possible to provide a more explicit expression for the incentive rates. Indeed, the first-order conditions provide the optimal incentive payment rate $z_i$:
\begin{align*}
0=& \frac{q_{-i}}{\Qm} (v_{x_i} - l_i)  -\frac{\eps}{\Qm} v_{x_{-i}}
+ \frac{\eps}{\Qm} l_{-i}
- q_i \frac{q_{-i}}{\Qm} a_i
+ q_{-i} \frac{\eps}{\Qm} a_{-i} \\
& - \eps \Big( \frac{q_{-i}}{\Qm} a_{-i} - \frac{\eps}{\Qm} a_i(z)\Big) 
- |b_i|^2\big( \etaA z_i + \etaP(z_i - v_{x_i})\big). 
 \end{align*}
By direct calculation, this provides
\begin{align*}
z_i & = \Big( 1 - \etaA \frac{Q_{-i}(\gamma_{-i}) \Qm | b_i |^2}{\bQ(\gamma)}\Big) v_{x_i} 
- \eps \etaA \frac{\Qm | b_{-i}|^2}{\bQ(\gamma)} v_{x_{-i}},
\end{align*}
where we used the notations
$$
Q_i(\gamma_i) := q_{-i} + \Qm (\etaA+\etaP) |b_i(\gamma_i)|^2
~\mbox{and}~
\bQ(\gamma) := Q_i(\gamma_i) Q_{-i}(\gamma_{-i}) - \eps^2 > 0.
$$

\subsection{Regulator facing two competing producers}\label{app:twoprod}

We recall that the incentive scheme provided to firm $i$ is given by \eqref{eq:xi2}. Then we obtain the dynamics of the total payment process $Y := Y^1 + Y^2$ for the two incentive mechanisms:
\begin{align*}%\label{eq:x2}
\dd Y_t^{Z, \Gamma} & = 
    \big[g(\ac(Z_t)) + \phi(b(\Gamma_t)) - p (X^1_t+X^2_t) \big] \dd t \\
&  + \frac12  \big\{  \big[  \eta_1(Z^{1,1}_t)^2 + \eta_2(Z^{2,1}_t)^2\big] | b_1(\Gamma^1_t)|^2 
    +                        \big[ \eta_2(Z^{2,2}_t)^2 + \eta_1(Z^{1,2}_t)^2\big] | b_2(\Gamma^2_t)|^2 \big\} \dd t \\
  & +  (Z^{1,1}_t  + Z^{2,1}_t) b_1(\Gamma^1_t) \cdot \dd W^1_t + (Z^{2,2}_t + Z^{1,2}_t) b_2(\Gamma^2_t) \cdot \dd W^2_t.
\end{align*}

We turn now to the PDE satisfied by the value function $V(t,x,y)$ of the regulator. It holds that
\begin{align*}
-\partial_t V = \sup_{z,\gamma}\Big\{ &
\sum_{i=1,2} -\Big[ g_i(a(z)) + \phi_i(b_i(\gamma_i)) 
+  \frac12  (\etai z_{i,i}^2 + \etaj z_{j,i}^2) |b_i(\gamma_i)|^2  - px_i \Big] V_y 
+ (a_i(z) - \delta_i x_i) V_{x_i}\\
&+ \frac12 \big[z_{i,i} + z_{j,i}\big]^2 |b_i(\gamma_i)|^2  V_{yy} 
 + \frac12 |b_i(\gamma_i)|^2 V_{x_ix_i} 
 + \big(z_{i,i} + z_{j,i}\big) |b_i(\gamma_i)|^2 V_{yx_i} \\
& -\etaP\big[ (k_i-p)x_i - \frac12 h |b_i(\gamma_i)|^2\big] V 
\Big\}, \quad
V(T,x,y) = U_{\rm P}(-y).
\end{align*}
Letting $V(t,x,y) = U_{\rm P}(v(t,x) - y)$, we get
\begin{align*}
-\partial_t v = \sup_{z,\gamma}\Big\{ &
\sum_{i=1,2} - g_i(a(z)) - \phi_i(b_i(\gamma_i)) 
- \frac12   (  \etai z_{i,i}^2 +  \etaj z_{j,i}^2) |b_i(\gamma_i)|^2
+ (a_i(z) - \delta_i x_i) v_{x_i}  \\
& - \frac12 \etaP \big[z_{i,i}+z_{j,i}\big]^2 |b_i(\gamma_i)|^2  
 + \frac12 |b_i(\gamma_i)|^2 (v_{x_ix_i} - \etaP v_{x_i}^2) \\
& + \etaP \big[z_{i,i} + z_{j,i} \big]  |b_i(\gamma_i)|^2 v_{x_i}  
 + k_i x_i - \frac12 h |b_i(\gamma_i)|^2 
\Big\}, \quad
v(T,x) = 0.
\end{align*}
Thus, dropping implicit dependencies on $z$ and $\gamma$ in $a$ and $b$, the HJB rewrites
\begin{align*}
-\partial_t v =& \sum_{i=1,2} (k_i-\delta_i v_{x_i}) x_i 
 + \sup_{z,\gamma} \HcP(z,\gamma), \\ 
 \text{with} & \quad \HcP(z,\gamma) :=  \sum_{i=1,2}  a_i  v_{x_i} - g_i(a) - \phi_i(b_i)    
 + \frac12 |b_i |^2 m_i(z,v_{x_i},v_{x_ix_i}),  \\
\text{and} & \quad
m_i^{\rm c}(z,v_i,v_{ii}) := 
v_{ii} -  \etai z_{i,i}^2 - \etaj z_{j,i}^2 - \etaP (z_{i,i} + z_{j,i} - v_i)^2 - h.
\end{align*}
As in the single firm case, we optimise over $\gamma$ first and get a similar expression
\begin{align*}
\gamma_i = m_i^{\rm c}(z,v_{x_i},v_{x_ix_i}) \wedge 0, \quad i=1,2.
 \end{align*}

\hs

Now, optimization over $z =(z_{1,1}, z_{2,2}, z_{1,2}, z_{2,1})^\intercal$ resolves in finding the maximiser $z$ of the function
\begin{align*}
& H_{\rm p}(z,\gamma)  :=  Dv \cdot a^{\rm c}(z) - L \cdot a^{\rm c}(z) - \frac12 a^{\rm c}(z) \cdot G a^{\rm c}(z) - \frac12 z \cdot M_{\rm c}(\gamma) z -  N_{\rm c}(\gamma) \cdot z, \quad  \text{with}\\ 
 & a^{\rm c}(z) = A_1^{\rm c} z + A_0, \quad 
A_0 := - G_{\rm c}^{-1} L,
 \end{align*}
where $G$ and $L$ have been defined above in the single firm case in relation~\eqref{eq:G}, and $A_1^{\rm c}$, $G_1^{\rm c}$ have been defined in~\eqref{eq:A1c} and $M_{\rm c}$ and $N_{\rm c}$ have been defined in~\eqref{eq:Mc}. Besides, we have $Dv \cdot   A_1^{\rm c} z$    $= (A_1^{\rm c})^\intercal Dv \cdot z$, and
\begin{align*}
a(z) \cdot G a(z) & 
% =  A_1 z \cdot G A_1 z + A_1 z \cdot G A_0 + A_0 \cdot G A_1 z + A_0 \cdot G A_0 \\
 = z \cdot (A_1^{\rm c})^\tp G A_1^{\rm c} z +  2 (A_1^{\rm c})^\tp G A_0 \cdot z + A_0 \cdot G A_0.
\end{align*}
Hence, first-order condition wrt $z$ provides
\begin{align*}
(A_1^{\rm c})^\tp (Dv - L) -  (A_1^{\rm c})^\tp G A_1^{\rm c} z -  (A_1^{\rm c})^\tp G A_0 - M(\gamma) z - N(\gamma) =0, \\
\big(M_{\rm c}(\gamma) + (A_1^{\rm c})^\tp G A_1^{\rm c}\big) z 
= (A_1^{\rm c})^\tp Dv + (A_1^{\rm c})^\tp( G G_{\rm c}^{-1} -\mathbb I_2)L  - N_{\rm c}(\gamma). 
\end{align*}
Assuming $M_{\rm c} +  (A_1^{\rm c})^\tp G A_1^{\rm c}$ to be invertible, the incentive payment rates $z \in \R ^4$ and $\gamma \in \R^2$ are solutions of the system
\begin{align}\label{app:eq:zgsb-c}
 &
 z  = \big(M_{\rm c}(\gamma) +  (A_1^{\rm c})^\tp G A_1^{\rm c}\big)^{-1} \big\{ (A_1^{\rm c})^\tp \big(Dv   + \big[G G_{\rm c}^{-1} -\mathbb I_2\big]L\big)  - N_{\rm c}(\gamma) \big\}, \\
&\gamma_i  = m_i^{\rm c}(z,v_{x_i},v_{x_ix_i}) \wedge 0, \quad i=1,2. \nonumber
\end{align}
Now, making the same educated guess as in the single firm case that $v(t,x_1,x_2) = w_0(t) + w_1(t) x_1 + w_2(t) x_2$, we deduce that $- \dot w_i = k_i - \delta_i w_i, \quad w_i(T) = 0$, $i=1,2$ (the same as in the single firm case), and $-\dot w_0(t) = \HcP(z(t),\gamma(t)),$ $w_0(T) = 0$, where $z$, $\gamma$ and  $m$ are given by~\eqref{app:eq:zgsb-c}. This completes the derivation of a solution $V$ of the HJB equation which is then easily proved to coincide with the required value function by a standard verification argument that we do not report for simplicity.

\hs

Finally, the form of the contract for each firm given in Proposition~\ref{prop:twoprodsb}~(ii) deduces in the same way in the single firm case. Indeed, using the form~\eqref{eq:xi2} of the contrat for firm~$i$ and applying the same technic of integration by part as in the monopoly case, one gets that
\begin{align*}
\int_0^T z_{i,i}(t) \dd X^i_t - (p - \delta_i X^i_t) z_{i,i}(t) \dd t = (z_{i,i}(T) X^i_t - z_{i,i}(0) X^i_0)  + \int_0^T (\delta_i z_{i,i}(t) - p - \dot z_{i,i}(t)) (X^i_t -X^i_0) \dd t.
\end{align*}
Same computations provides the form for term of incentives $z_{i,j}(t) dX^j_t$. But in the case of two competiting firms, it may occur that $z_{,ii}(T)$ and $z_{i,j}(T)$ are non-zero. This explains the presence of the terminal payment in the contract form \eqref{eq:main:xic}. Regarding the incentive price $\ricii$, it appears when using the equality $\gamma_i(t) + \etai z_{i,i}^2$ $=$ $-\etaj z_{j,i}^2 - \etaP (z_{i,i} + z_{j,i} - w_i(t))^2$ while $-\ricij$ is simply $\eta_i z_{j,i}^2$ for $j\neq i$.

\subsection{Market structures comparison}\label{app:compare}

We compare here the certainty equivalent of the regulator facing a single firm handling the two technologies and a regulator facing two competing firms, having the same risk-aversion $\etaA = \eta_1 = \eta_2 =: \eta$. We denote here $v^{\rm m}$ and $v^{\rm c}$ the certainty equivalent in case of a single firm and in case of two competing firms respectively. We have
\begin{align*}
& v^{\rm m}(t,x_1,x_2) = w_0^{\rm m}(t) + w_1(t) x_1 + w_2(t) x_2, \quad
v^{\rm c}(t,x_1,x_2) = w_0^{\rm c}(t) + w_1(t) x_1 + w_2(t) x_2,  \\
& \text{with}  \quad
 w_0^{\rm m}(t) = \int_t^T \HcP^{\rm m}(\zm(t),\gm(t)) \dd t,  \quad
 w_0^{\rm c}(t) = \int_t^T \HcP^{\rm c}(\zc(t),\gc(t)) \dd t, \\
 & 
 \HcP^{\rm m}(\zm(t),\gm(t)) = \sup_{z \in \R^2,\gamma \in \R^2} \Big\{
\sum_{i=1,2}  a_i(z)  w_{i}(t) - g_i(a(z)) - \phi_i(b_i)    
 + \frac12 |b_i |^2 m_i(z,w_i(t)) \Big\}, \\
 & \HcP^{\rm c}(\zc(t),\gc(t)) = \sup_{z \in \R^4, \gamma \in \R^2} \Big\{
 \sum_{i=1,2}  a_i^\ast(z)  w_i(t) - g_i(\ac(z)) - \phi_i(b_i)    
 + \frac12 |b_i |^2 m_i^{\rm c}(z,w_i(t)) \Big\},  \\
 & \quad m_i(t,z) :=  -  \eta  z_i^2 - \etaP (z_i - w_i(t))^2  - h, \quad
m_i^{\rm c}(t,z) := 
 - \eta (z_{i,i}^2 + z_{j,i}^2) - \etaP (z_{i,i} + z_{j,i} - w_i(t))^2 - h
 \end{align*}
where $(\zm(t), \gm(t))$ and $(\zc(t), \gc(t))$ are solutions of \eqref{eq:zgsb-z}-\eqref{eq:zgsb-g} and \eqref{eq:zgsb-c} respectively.	Showing $v^{\rm m}(t,x_1,x_2) \le v^{\rm c}(t,x_1,x_2)$ is equivalent to showing $w_0^{\rm m}(t) \le w_0^{\rm c}(t)$. Thus, it is enough to show that $\HcP^{\rm m}(\zm(t),\gm(t)) \le \HcP^{\rm c}(\zc(t),\gc(t))$.

\hs

Now, observe that for any $z := (z_1,z_2) \in \R^2$, there exists a vector $\bar z := (z_{1,1}, z_{2,2}, z_{1,2}, z_{2,1}) \in \R^4$, such that
\begin{align*}
&a_i(z) =: u_i = \ac_i(\bar z), \quad
 m_i(t,z) = m_i^{\rm c}(t,\bar z), \quad i=1,2.
\end{align*}

Indeed, note that when  $z_{i,i} = z_i$, and $z_{i,j} = 0$, for $j\neq i$, it holds that $m_i^{\rm c}(t,\bar z) =  m_i(t,z)$ =$\gamma_i$. Now, taking null cross-payments $z_{i,j}$, we are left with a linear system
\begin{align*}
& u_i =  \ac_i(\bar z), \quad i=1,2
\end{align*}
with non-zero determinant $Q_{\rm c} = q_1q_2- \frac{\eps^2}{4}$. Thus, there exists a solution $(z_{1,2}, z_{2,1})$ for all $(u_1,u_2)$. Hence, $\HcP^{\rm m}(\zm(t),\gm(t)) \leq \HcP^{\rm c}(\zc(t),\gc(t))$. The result deduces directly. \hfill$\Box$

\end{document}